\documentstyle[aps,twocolumn,epsf,psfig]{revtex}

\begin{document}
\newtheorem{theorem}{Theorem}
\newtheorem{acknowledgement}[theorem]{Acknowledgement}
\newtheorem{algorithm}[theorem]{Algorithm}
\newtheorem{axiom}[theorem]{Axiom}
\newtheorem{claim}[theorem]{Claim}
\newtheorem{conclusion}[theorem]{Conclusion}
\newtheorem{condition}[theorem]{Condition}
\newtheorem{conjecture}[theorem]{Conjecture}
\newtheorem{corollary}[theorem]{Corollary}
\newtheorem{criterion}[theorem]{Criterion}
\newtheorem{definition}[theorem]{Definition}
\newtheorem{example}[theorem]{Example}
\newtheorem{exercise}[theorem]{Exercise}
\newtheorem{lemma}[theorem]{Lemma}
\newtheorem{notation}[theorem]{Notation}
\newtheorem{problem}[theorem]{Problem}
\newtheorem{proposition}[theorem]{Proposition}
\newtheorem{remark}[theorem]{Remark}
\newtheorem{solution}[theorem]{Solution}
\newtheorem{summary}[theorem]{Summary}    
\def\r{{\bf{r}}}
\def\i{{\bf{i}}}
\def\j{{\bf{j}}}
\def\m{{\bf{m}}}
\def\k{{\bf{k}}}
\def\kt{{\tilde{\k}}}
\def\mt{{\hat{t}}}
\def\mG{{\hat{G}}}
\def\mg{{\hat{g}}}
\def\mGa{{\hat{\Gamma}}}
\def\mS{{\hat{\Sigma}}}
\def\mT{{\hat{T}}}
\def\K{{\bf{K}}}
\def\P{{\bf{P}}}
\def\q{{\bf{q}}}
\def\Q{{\bf{Q}}}
\def\p{{\bf{p}}}
\def\x{{\bf{x}}}
\def\y{{\bf{y}}}
\def\X{{\bf{X}}}
\def\Y{{\bf{Y}}}
\def\F{{\bf{F}}}
\def\G{{\bf{G}}}
\def\bG{{\bar{G}}}
\def\bV{{\bar{V}}}
\def\mbG{{\hat{\bar{G}}}}
\def\M{{\bf{M}}}
\def\V{\cal V}
\def\tchi{\tilde{\chi}}
\def\tx{\tilde{\bf{x}}}
\def\tk{\tilde{\bf{k}}}
\def\tK{\tilde{\bf{K}}}
\def\tq{\tilde{\bf{q}}}
\def\tQ{\tilde{\bf{Q}}}
\def\si{\sigma}
\def\ep{\epsilon}
\def\hep{{\hat{\epsilon}}}
\def\al{\alpha}
\def\be{\beta}
\def\ep{\epsilon}
\def\bep{\bar{\epsilon}_\K}
\def\up{\uparrow}
\def\de{\delta}
\def\De{\Delta}
\def\up{\uparrow}
\def\dwn{\downarrow}
\def\ksi{\xi}
\def\etha{\eta}
\def\product{\prod}
\def\goto{\rightarrow}
\def\switch{\leftrightarrow}
                           
\title{Dynamical Cluster Approximation Employing FLEX as a Cluster Solver} 
\author{
K. Aryanpour$^{1}$
\and	
M. H. Hettler$^{2}$ and M. Jarrell$^{1}$ 
}
\address{$^{1}$University of Cincinnati, Cincinnati OH 45221, USA}
\address{$^{2}$Forschungszentrum Karlsruhe, Institut f\"ur Nanotechnologie,
 Karlsruhe, Germany}

\maketitle

\begin{abstract}
We employ the Dynamical Cluster Approximation (DCA) in conjunction with 
the Fluctuation Exchange Approximation (FLEX) to study the Hubbard
model. The DCA is a technique to systematically restore the momentum 
conservation at the internal vertices of Feynman diagrams  
relinquished in the Dynamical Mean Field Approximation (DMFA). FLEX 
is a perturbative diagrammatic approach in which classes of 
Feynman diagrams are summed over analytically using 
geometric series. The FLEX is used as a tool to 
investigate the complementarity of the DCA and the finite size 
lattice technique with periodic boundary conditions by comparing their 
results for the Hubbard model. We also study the microscopic theory underlying 
the DCA in terms of compact (skeletal) and non-compact diagrammatic 
contributions to the thermodynamic potential independent of a specific model. 
The significant advantages of the DCA implementation in momentum space 
suggests the development of the same formalism for the frequency space. 
However, we show that such a formalism for the Matsubara frequencies at 
finite temperatures leads to acausal results and is not viable. However, 
a real frequency approach is shown to be feasible.  
\end{abstract}

\section {Introduction}
Non-local correlations play an important role in the physics of 
strongly correlated electron systems such as high-$Tc$ superconductors, 
heavy fermion metals, etc. The Dynamical Mean Field Approximation
(DMFA) \cite{Pruschke,Georges}, in which all the non-local correlations 
are ignored, can capture some of the major features of strongly 
correlated systems. Nevertheless, the non-local correlations become 
crucial in the physics of phases with non-local order parameters such 
as d-wave superconductivity. Even phases with local order parameters 
such as commensurate magnetism can be significantly affected by the 
non-local correlations (e.g.\ spin waves) ignored in the DMFA.

The early attempts to extend the DMFA by including non-local 
correlations resulted in the violation of causality which is a 
requirement for positive definiteness of the spectral weight and 
the density of states (DOS) \cite{van Dongen}. The Dynamical Cluster 
Approximation (DCA) is a fully causal technique used to systematically 
add nonlocal corrections to the DMFA by mapping the lattice onto a 
self-consistently embedded cluster problem. The mapping from the 
lattice to the cluster is accompanied by coarse-graining the lattice 
problem in its reciprocal space. Thus far, the DCA has been combined 
with Quantum Monte Carlo (QMC) \cite{DCA_hettler,DCA_hettler2}, the 
Non-Crossing Approximation (NCA) \cite{DCA_maier1} and the Fluctuation 
Exchange Approximation (FLEX) \cite{Aryanpour,Imai} to solve the 
corresponding cluster problems. 

The FLEX is a perturbative diagrammatic approach in which classes 
of Feynman diagrams are summed to all orders using geometric 
series.\cite{Bickers1,Bickers2}  Others \cite{Deisz,Serene} have 
employed the FLEX for finite size lattices with periodic boundary conditions. 
Due to the absence of contributions from some relevant diagrams, the FLEX 
is not capable of addressing the Hubbard model physics in the strong regime 
precisely. However, the main objective of this work is to make a comparison 
between the DCA-FLEX combination results and previous finite size 
lattice FLEX calculations. It is hoped that this study will lead to 
a better understanding of the DCA.
    
We earlier\cite{Aryanpour} suggested a prescription to correctly 
implement the DCA technique in the thermodynamic potential. This 
prescription will be discussed from a different point of view using a 
more general argument. Based upon the Green function's exponential
fall-off as a function of distance, we conclude that compact diagrams 
(two-particle irreducible in the thermodynamic potential) are better 
approximated using the DCA than non-compact (two-particle reducible) 
ones. Hence, the DCA is applied to the compact diagrams only and
non-compact ones are calculated explicitly using dressed non
coarse-grained Green functions.

In this work we also consider the extension of the DCA to frequency 
space. The many-body theory 
at finite temperatures is conventionally derived in terms of discrete 
imaginary Matsubara frequencies. We illustrate that even for 
a self-consistent algorithm like the FLEX, coarse-graining the
imaginary frequency propagators results in causality violations
and can not be implemented. However, a real frequency formalism is
shown to be causal and applicable not only to the FLEX, but also to 
other cluster solving methods such as the NCA.        

This paper is structured as follows.  In the next three sections, we 
briefly review the DCA, its application to the Hubbard model, the
FLEX, and then we describe how the FLEX and the DCA may be merged
into a single algorithm.  In the next three sections, we use the 
FLEX-DCA, in comparison to the FLEX for finite-sized systems, to
explore the properties of the DCA.  The last two sections, are devoted 
to a microscopic derivation of the DCA, and to an extension of the
DCA to frequency space.

\section{Dynamical Cluster Approximation (DCA)}
\label{Dynamical}
\begin{figure}
\epsfxsize=1.8in
\hspace{1.5cm}\epsffile{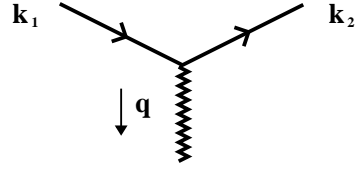}
\vspace{0.5cm}
\caption[a]{\em A typical vertex in a Feynman diagram with solid 
lines as one particle Green functions and the wiggly line as an interaction.}
\label{Laue}
\end{figure}

Both the DCA and the DMFA may be derived by exploring the momentum
conservation in the diagrammatics.  As depicted in Fig.~\ref{Laue}, 
momentum conservation at each vertex is described by the Laue function:
\vspace{0.1cm}
\begin{eqnarray}
\label{eq:Laue}
\Delta=\sum_\x e^{i\x\cdot(\k_1-\q-\k_2)} = N\delta_{\k_1,\q+\k_2}\,,
\end{eqnarray}
In the DMFA, momentum conservation at the internal vertices of 
irreducible Feynman diagrams is completely relinquished.  I.e., the 
DMFA simply sets $\De=1$. \cite{muller-hartmann} Hence, we may sum 
freely over all the internal momenta entering and leaving each vertex.
Only local contributions survive the sum. Thus, this is equivalent to
mapping  the lattice problem onto a self-consistently embedded
impurity problem. The DMFA becomes exact at infinite dimensions. 
\cite{Metzner}

\begin{figure}
\epsfxsize=3.3in
\epsffile{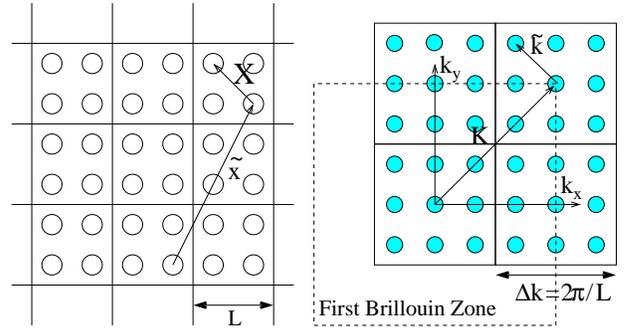}
\vspace{0.1cm}
\caption[a]{\em{The real lattice clusters (right) and (left) the first
 Brillouin zone divided into subcells.}}
\label{divide_x_k}
\end{figure}

The DCA is an approach to systematically restore the momentum 
conservation relinquished in the DMFA. In the DCA, the first 
Brillouin zone in the reciprocal space is divided into $N_c$ 
equal cells of linear size $\Delta k$ labeled by $\K$ in their 
centers, and the momenta within each cell are labeled by $\tk$.
Then $\k=\K+\tk$ (c.f. Fig.~\ref{divide_x_k}). To visualize this 
scheme in the real lattice, one could consider tiling the lattice 
of $N$ sites by $N/N_{c}$ clusters each composed of $N_{c}=L^D$ 
sites where $L$ is the linear size of the subcell and D is 
dimensionality (c.f. Fig.~\ref{divide_x_k} for $L=2$). We will 
use this picture in section \ref{microscopic} while discussing 
the microscopic theory of the DCA. We label the origin of the 
clusters by $\tx$ and the $N_{c}$ intercluster sites by $\X$. 
So for each site in the original lattice $\x=\X+\tx$. Care must 
be taken when choosing the cluster geometries in order to 
preserve the lattice point group symmetry and also satisfy 
some other criteria for cubic or square lattices. \cite{Betts} 

In the DCA, we first make the following separation in Eq.~\ref{Laue}
\begin{eqnarray}
\label{eq:LSeparationr}
\Delta=\sum_\x e^{i(\tx+\X)\cdot(\K_1-\Q-\K_2+\tk_1-\tq-\tk_2)}\,.
\end{eqnarray}
The products $\K_1\cdot\tx, \Q\cdot\tx$ and $ \K_2\cdot\tx = 2n\pi$ 
where $n$ is an integer. Therefore, their associated phases may be
neglected and Eq.~\ref{Laue} splits into two parts 
\begin{equation}
\label{eq:LDCA1}
\Delta=\frac{N}{N_c}\hspace{0.1cm}\de_{\tk_1,\tq+\tk_2}\hspace{0.1cm}
N_c\hspace{0.1cm}\de_{\K_1,\Q+\K_2}\,.
\end{equation}
The DCA also ignores the phases $e^{-i\tk\cdot\tx}$ due to the
position of the cluster in the original lattice and (far less important) 
$e^{-i\tk\cdot\X}$ corresponding to the position within the cluster. 
As a result, it approximates 
$N/N_c\hspace{0.1cm}\de_{\tk_1,\tq+\tk_2}\cong 1$, so that
\begin{equation}
\label{eq:LDCA2}
\Delta_{DCA}=N_c\hspace{0.1cm}\de_{\K_1,\Q+\K_2}\,,
\end{equation}
which indicates that the momentum is conserved modulo $\Delta k$ 
for transfers between the cells. 
\vspace{0.1cm}

The approximation made through the substitution $\De \to \De_{DCA}$ 
corresponds to replacing all internal legs in the compact (skeletal) diagrams 
by the coarse-grained Green function $\bG$ and interaction potential
$\bV$  defined by
\begin{equation}
\label{eq:cgGDCA}
\bG(\K,\omega_n)=\frac{N_c}{N}\sum_{\tk}G(\K+\tk,\omega_n)\,,
\end{equation} 
\begin{equation}
\label{eq:cgPDCA}
\bV(\Q)=\frac{N_c}{N}\sum_{\tq}V(\Q+\tq)\,.
\end{equation} 
In section~\ref{microscopic}, we will define the compact and non
compact diagrams and elaborately discuss why only the compact ones 
undergo the coarse-graining approximation.

Replacing $\Delta$ by $\Delta_{DCA}$ tremendously reduces the
complexity of the problem because instead of having to perform 
sums over all $N$ states in the entire first Brillouin zone, 
we have sums over only a set of $N_{c}$ states where $N_{c} << N$.
\vspace{-0.2cm}
\section{Hubbard Model}
We will apply the DCA to study the Hubbard model Hamiltonian incorporating 
interactions between the electrons on a lattice. It includes a tight-binding 
part due to the hopping of electrons among the sites and an interaction 
between the electrons. The general Hamiltonian reads
\begin{equation}
\label{Hubbard1}
H = H_{0} + H_{I}\,,
\end{equation} 
where
\begin{equation}
\label{Hubbard2}
H_{0} = -t\sum_{\sigma}\sum_{<ij>}c^{\dagger}_{i\sigma}c_{j\sigma}\,,
\end{equation} 
 and 
\begin{eqnarray}
\label{Hubbard3}
H_{I} =\frac{1}{2}\sum_{jl,\sigma\sigma'}n_{j\sigma}n_{l\sigma'}
V(R_{j}-R_{l})\nonumber\\&&\hspace{-5.0cm}V(R)\approx \frac{e^2}{R}\,.
\end{eqnarray} 
Factors $t$ and $V(R)$ correspond to electron hoppings and Coulomb 
interactions respectively. Later in the paper, we will study the simplest 
Hubbard interaction which is fully local and only between electrons 
sitting at the same site having opposite spin directions. The interaction 
strength is a constant called U. Hence, for the local model, 
Eq.~\ref{Hubbard3} simplifies to    
\begin{equation}
\label{Hubbard4}
H_{I} = U\sum_{i}n_{i\uparrow}n_{i\downarrow}\,.
\end{equation} 
In terms of the vertex properties addressed in section~\ref{Dynamical},
since the interaction is local and therefore independent of $\q$, 
we may sum freely over the $\q$ momentum for a pair of Laue functions
in Eq.~\ref{Laue} sharing a common interaction wiggly line as
depicted in Fig.~\ref{Laue}. As a result, the  corresponding Laue 
function will become
\vspace{0.1cm}
\begin{eqnarray}
\label{Laue2}
\Delta=\frac{1}{N}\sum_\q\sum_\x e^{i\x\cdot(\k_1-\q-\k_2)}
\sum_\y e^{i\y\cdot(\k_3+\q-\k_4)}=\nonumber\\&&\hspace{-7.0cm}N
\sum_\q\de_{\k_1,\q+\k_2}~\de_{\k_3+\q,\k_4}=N\de_{\k_1+\k_3,\k_2+\k_4}\,,
\end{eqnarray} 
and analogously for the DCA, by summing freely over $\Q$
\begin{equation}
\label{Laue3}  
\Delta_{DCA}=N_c~\de_{\K_1+\K_3,\K_2+\K_4}\,.
\end{equation} 

\section{Fluctuation Exchange Approximation (FLEX)}
\label{Fluct-App}
\begin{figure}
\epsfxsize=2.7in
\hspace{-0.3cm}\epsffile{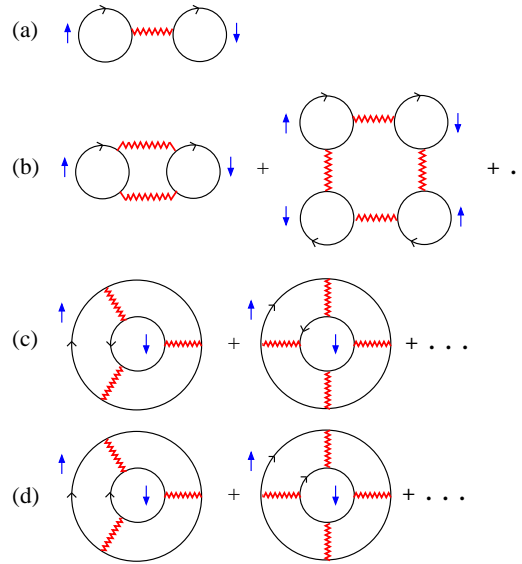}
\vspace{0.4cm}
\caption[a]{
(a) Lowest order diagram (Hartree term), 
(b) longitudinal spin and density fluctuation $\Phi$  
diagrams with an even number of bubbles ($\Phi_{ph}^{df}$), 
(c) transverse spin fluctuations ($\Phi_{ph}^{sf}$) and 
(d) particle-particle fluctuations ($\Phi_{pp}$).} 
\label{FLEX}
\end{figure}

In the Feynman diagrammatics of the Hubbard model with a local 
interaction, all the interactions (wiggly lines in Fig.~\ref{FLEX}) 
contribute a c-number U from Eq.~\ref{Hubbard4}. The electronic Green 
functions (solid lines)  which interact with one another should 
have opposite spins. Considering these restrictions, had we 
been able to include all the possible diagrams in our expansion 
we would have solved the problem exactly. However, in practice 
this is not feasible.

The Fluctuation Exchange Approximation (FLEX) was introduced as an 
approximate technique to simplify this diagrammatic sum 
\cite{Bickers1,Bickers2}, while retaining a conserving approximation. 
In the FLEX, the interaction part of the Hubbard model Hamiltonian 
is treated perturbatively by selecting a certain class of all the 
possible diagrams which may be summed as a geometric series. Following 
Baym \cite{Baym}, we define the generating functional $\Phi[G(k,\omega_n)]$ 
as the collection of all the selected families of diagrams illustrated in 
Fig.~\ref{FLEX}. Therefore, $\Phi[G(k,\omega_n)]$ for the FLEX 
can be written
\begin{equation}
\label{eq:FLEX_phi}
\Phi=\Phi_{ph}^{df}+\Phi_{ph}^{sf}+\Phi_{pp}\,,
\end{equation} 
\begin{eqnarray}
\label{FLEX_phi_a}
\hspace{-0.4cm}\Phi_{ph}^{df} =  -\frac{1}{2}\hspace{0.1cm}\rm{Tr}\hspace
{0.1cm}[\chi_{ph}]^2+\frac{1}{2}\hspace{0.1cm}\rm{Tr}\hspace{0.1cm}[\rm{ln}
(1+\chi_{ph})-\nonumber\\&&\hspace{-6.5cm}\chi_{ph}+\frac{1}{2}
{\chi^2}_{ph}]\,,
\end{eqnarray}
\begin{equation}
\label{eq:FLEX_phi_b}
\Phi_{ph}^{sf} = \frac{3}{2}\hspace{0.1cm}\rm{Tr}\hspace{0.1cm}[\rm{ln}
(1-\chi_{ph})+\chi_{ph}+\frac{1}{2}{\chi^2}_{ph}]\,,
\end{equation} 
\begin{equation}
\label{eq:FLEX_phi_c}
\hspace{-0.4cm}\Phi_{pp} = \rm{Tr}\hspace{0.1cm}[\rm{ln}(1+\chi_{pp})-
\chi_{pp}+\frac{1}{2}{\chi^2}_{pp}]\,,
\end{equation} \\
where Tr = $(T/N)\sum_{k}\sum_{n}$ with $T$ the temperature and $N$ the
number of lattice sites. The particle-hole and particle-particle
susceptibility bubbles are
\begin{eqnarray}
\chi_{pp}(q,\omega_{n})=\nonumber\\&&\hspace{-1.5cm}
U(T/N)\sum_{k}\sum_{m}G(k+q,\omega_{n}+\omega_{m})G(-k,-\omega_{m})\,,
\label{chipp}
\end{eqnarray}
\begin{eqnarray}
\chi_{ph}(q,\omega_{n})=\nonumber\\&&\hspace{-1.5cm}
-U(T/N)\sum_{k}\sum_{m}G(k+q,\omega_{n}+\omega_{m})G(k,\omega_{m})\,,
\label{chiph}
\end{eqnarray}
\vspace{0.2cm}
\vspace{0.2cm}
The self-energy and the Green function are defined by 
\begin{equation}
\label{eq:slfeng}
\Sigma(k,\omega_{n})= \frac{1}{2}\frac{\delta\Phi[G]}{\delta
G(k,\omega_{n})}\,,
\end{equation} 
\begin{equation}
\label{eq:greenfn}
G(k,\omega_{n})= \big[G^{(0)-1}(k,\omega_{n})-\Sigma(k,\omega_{n})
\big]^{-1}\,,
\end{equation} 
where $G^{(0)}$ is the non-interacting one particle Green function defined by
\begin{equation}
G^{(0)}(\k,\omega_{n}) =\frac{1}{\omega_{n}-\ep_\k+\mu} \,,
\label{G_bare}
\end{equation}
with $\ep_\k$ the non-interacting Hubbard model dispersion and $\mu$
 the chemical potential.

Calculating the self-energy for Eq.~\ref{eq:FLEX_phi} using 
Eq.~\ref{eq:slfeng} we get
\begin{eqnarray}
\label{eq:slfpot}
\Sigma(k,\omega_{n})= U(T/N)\sum_{q}\sum_{m}
\big[V^{(ph)}(q,\omega_{m})\times \nonumber\\
&&\hspace{-6.7cm}G(k-q,\omega_{n}-\omega_{m}) -
  V^{(pp)}(q,\omega_{m})\times\nonumber\\
&&\hspace{-6.7cm}G(-k+q,-\omega_{n}+\omega_{m})\big]\,,
\end{eqnarray}  
in which
\begin{eqnarray}
\label{eq:flex-pots1}
V^{(ph)}(q,\omega_{m}) = \chi_{ph}(q,\omega_{m})+
\frac{1}{2}\chi_{ph}(q,\omega_{m})\times\nonumber\\
&&\hspace{-6.8cm}\big[\frac{1}{1 + \chi_{ph}(q,\omega_{m})} - 1\big]
+\frac{3}{2}\chi_{ph}(q,\omega_{m}))\times\nonumber\\
&&\hspace{-6.8cm}\big[\frac{1}{1 - \chi_{ph}(q,\omega_{m})} - 1\big]\,,
\end{eqnarray}
\begin{equation}
\label{eq:flex-pots2}
V^{(pp)}(q,\omega_{m}) = \chi_{pp}(q,\omega_{m})
\big[\frac{1}{1 + \chi_{pp}(q,\omega_{m})} - 1\big]\,.
\end{equation}\\ 
Eq.~\ref{eq:flex-pots1},\ref{eq:flex-pots2} for the potential 
functions $V^{(ph)}$ and $V^{(pp)}$ are geometric series for 
$\chi_{ph}$ and $\chi_{pp}$ similar to the 
{\it random phase approximation} (RPA) results. The Hartree 
term contribution to the self-energy has not explicitly 
appeared in Eq.~\ref{eq:slfpot} as it is constant and can 
be always embedded in the chemical potential in Eq.~\ref{G_bare}.

The difference $\Delta\Omega(T,\mu)$ between interacting and 
non-interacting thermodynamic potential functional is also 
expressible in terms of the Green functions, self-energy, and $\Phi[G]$ \\ 
\begin{equation}
\label{eq:thermPot}
\Delta\Omega(T,\mu)=\Omega-\Omega_0 = -2 {\rm Tr} [\Sigma G - {\rm ln}
(G/G_{0})] + \Phi[G]\,.
\end{equation}

In the FLEX, since we include only a limited set of all the diagrammatic
contributions, we do not anticipate to precisely address the 
Hubbard model physics. However, there are a number of significant 
physical features such as anti-ferromagnetic order at half filling 
and low temperatures that this approximation is able to 
capture. Moreover, by using the FLEX both together with the DCA
and to study finite sized systems with periodic boundary conditions, we can study the 
differences between these approaches. For example, as we will shortly illustrate, 
the complementarity of the DCA and finite size lattice techniques is 
manifest in the FLEX. The FLEX can also be invoked as a good test 
for the microscopic theory of the DCA and the coarse-graining effects 
in the compact and non-compact diagrams for Eq.~\ref{eq:thermPot}.    
\vspace{-0.1cm}
\section{The Combination of the FLEX and DCA (Algorithm)}

\begin{figure}
\epsfxsize=2.3in
\hspace{1.0cm}\epsffile{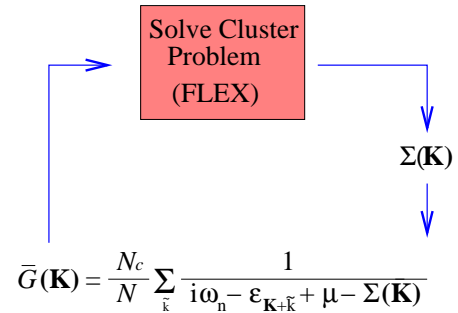}
\vspace{0.5cm}
\caption[a]{The FLEX-DCA numerical algorithm.\hspace{-0.05cm} 
The dressed $G$ is solved self-consistently with $\Sigma$. 
The iteration process stops whenever convergence is achieved.
}
\label{dca_flx_algrtm}
\end{figure}

In the combination of the FLEX and DCA, our goal is to calculate 
the self-energy in Eq.~\ref{eq:slfeng} whereby we construct the 
dressed Green function for the lattice as a building block for 
all the relevant physical quantities. We start out with the bare 
(non-interacting) Green function $G^{(0)}(\k,z)$ defined in 
Eq.~\ref{G_bare} with z the Matsubara frequency (complex). 
We coarse-grain $G$ as directed in Eq.~\ref{eq:cgGDCA} and 
calculate the self-energy using Eq.~\ref{eq:slfeng}. This is used to 
recompute the dressed Green function 
\begin{equation}
G(\k,z) =\frac{1}{z-\ep_\k+\mu-\Sigma_{DCA}(\K,z) } \,.
\label{G_DCA}
\end{equation} 
where the index DCA in $\Sigma_{DCA}(\K,z)$ indicates that we 
have used coarse-grained $\bG$ for the construction of self-energy. 
The new $G$ is 
coarse-grained and used to calculate a new estimate of  
$\Sigma_{DCA}(\K,z)$. We repeat this process iteratively 
until convergence at a desired tolerance is obtained. 
The final self-energy is used to construct the dressed Green function 
in Eq.~\ref{G_DCA}, required to compute the physical quantities such 
as spectral function, the density of states (DOS), etc. 
The algorithm of this calculation is demonstrated in Fig~\ref{dca_flx_algrtm}.
\section{Complementarity of the DCA to the Finite Size Lattice Approximation with periodic 
boundary conditions}
\begin{figure}
\epsfxsize=3.5in
\hspace{1.0cm}\epsffile{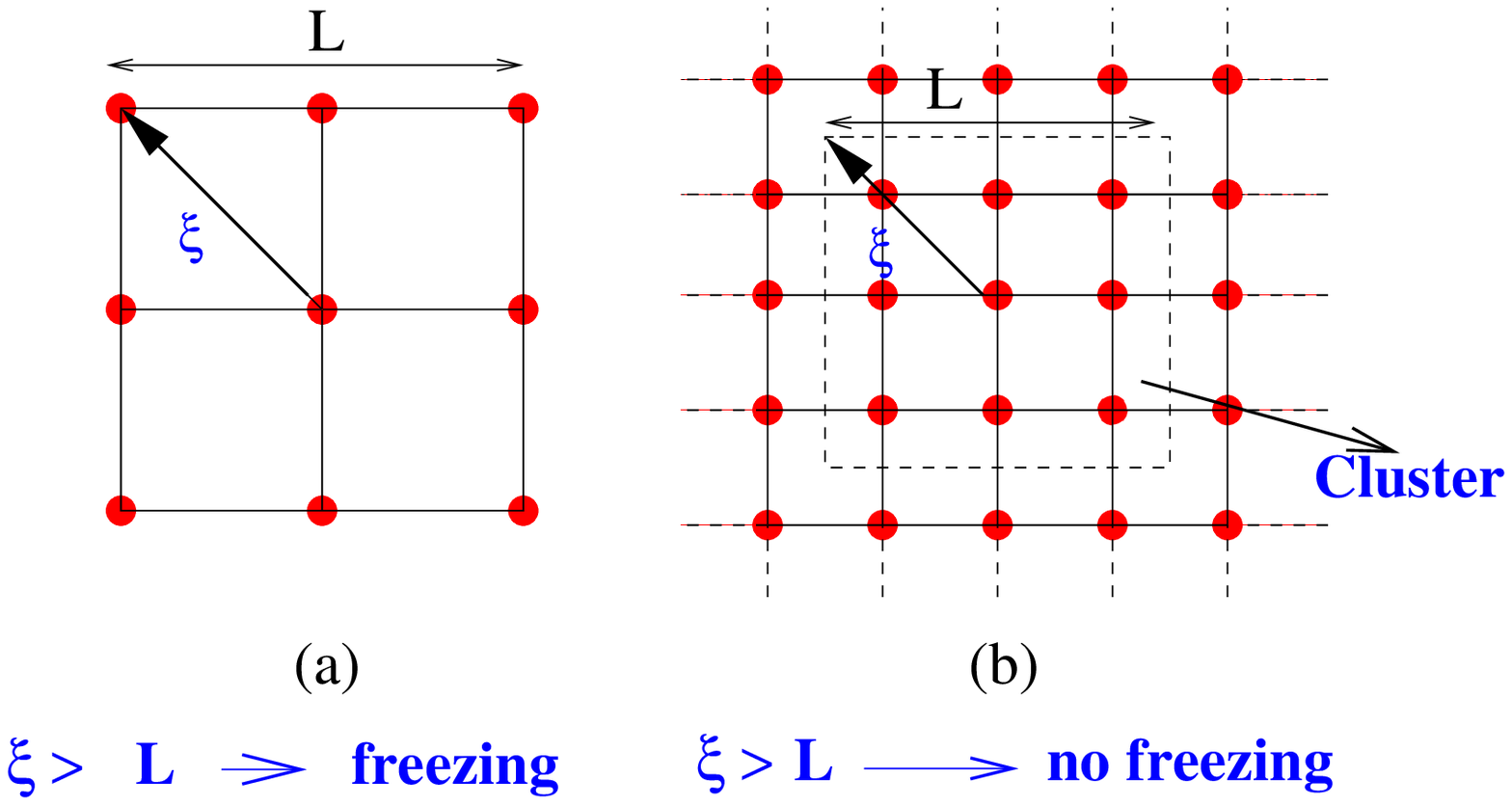}
\vspace{0.2cm}
\caption[a]{(a) The finite size lattice with periodic boundary conditions, size $L$ and correlation 
length $\xi$ and (b) the lattice with clusters of size $L$ and the correlation length $\xi$.}
\label{finvsDCA}
\end{figure}

In the half-filled Hubbard model, the antiferromagnetic correlation length
$\xi$ increases with decreasing temperature and diverges at the phase
transition. In a finite size lattice with periodic boundary conditions, as the temperature drops, 
once the correlation length reaches the size of the lattice, the system 
is fully frozen and there is a gap to excitations (c.f. Fig~\ref{finvsDCA}.a). 
In contrast, in the DCA, the correlations are confined within 
clusters of size $N_c << N$ (the size of the entire lattice) 
and they never reach the size of the lattice. As we lower the 
temperature, the correlation length approaches the size of the 
cluster but since the lattice remains in the thermodynamic limit, 
it never freezes (c.f. Fig~\ref{finvsDCA}.b). By increasing 
the size of clusters in the DCA, we take longer ranged correlations 
into account so the gap will become more pronounced. Consequently, 
correlation induced gaps are generally overestimated in the 
finite size lattice, while in the DCA they are underestimated.
\vspace{-0.5cm}
\vspace{-0.2cm}
\vspace{-0.2cm}
\begin{figure}
\epsfxsize=3.4in
\hspace{-0.5cm}\epsffile{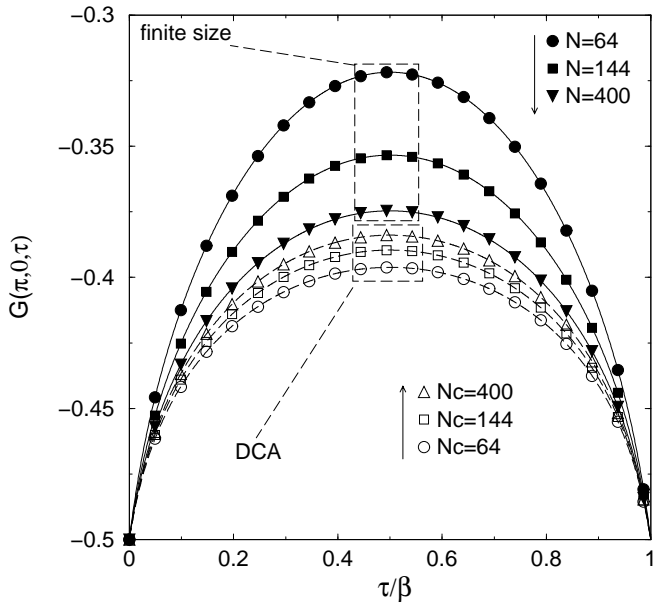}
\vspace{0.55cm}
\caption[a]{The Green function $G(k=(0,\pi),\tau)$ at $T=0.10$ and
$U/t=1.57$ versus imaginary time $\tau/\beta$  ($\beta=1/T$). 
Finite size results (filled symbols) and  DCA results (open symbols) 
are displayed for lattice sizes of $N=8\times8$, $12\times12$, and 
$20\times20$. 
The curves show the complementary approach of the two methods to 
the thermodynamic limit.
}
\label{dca_fnt_size}
\end{figure}
This complementary behavior may be seen in $G(k,\tau)$, with $k=(0,\pi)$
computed using finite size lattices with periodic boundary conditions and the DCA. As illustrated in Fig.~\ref{dca_fnt_size}, by increasing the size of the 
finite size lattice and the DCA cluster, the Green functions converge 
from opposite directions. In the finite size lattice, the Green function 
(e.g. at $\tau=\beta/2$) decreases with the increase of size which is 
consistent with overestimating the gap; while in the DCA, the 
Green function increases as the cluster size grows consistent 
with underestimating the gap. It is also observed that the 
convergence in the DCA is much faster,
meaning that the result of the DCA is closer to the true curve at a given 
cluster size.
Both finite size lattices with periodic boundary conditions and the DCA converge with corrections of 
${\cal O}(\lambda/L^2)$ 
with $L$ the linear size of the finite lattice or the DCA cluster and 
$\lambda$ a coefficient. \cite{DCA_maier2} The faster convergence of 
the DCA corresponds to its smaller $\lambda$ compared to 
finite size lattices.
\vspace{-1.8cm}
\begin{figure}[tbp]
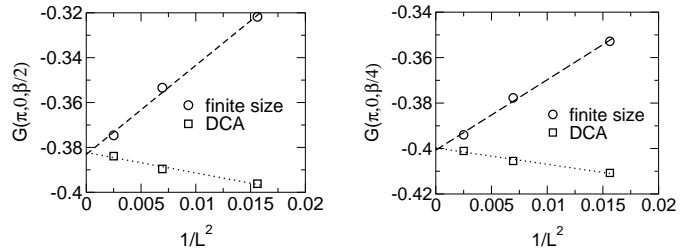

\begin{picture}(0,145)(10,0)
\put(0,0){\psfig{figure=./extrpl1.eps,height=1.27in}}
\end{picture}
\begin{picture}(0,145)(10,0)
\put(130,0){\psfig{figure=./extrpl2.eps,height=1.27in}}
\end{picture}
\vspace{0.25cm}
\caption[a]{The Green function $G(k=(0,\pi),\tau)$  at $\tau=\beta/2$
(left) and $\tau=\beta/4)$ (right) versus $1/L^2$ for the DCA and
finite size results in Fig.~\ref{dca_fnt_size}. The linear
extrapolations meet at a point representing the thermodynamic limit
of the Green function of the Hubbard model evaluated within FLEX.
}
\label{extprl}
\end{figure}
In Fig.~\ref{extprl}, the values of the $G(k=(0,\pi),\tau=\beta/2)$ and 
$G(k=(0,\pi),\tau=\beta/4$ in Fig.~\ref{dca_fnt_size} have been plotted 
versus $1/L^2$ for both the DCA and finite size results. The Green function 
behaves linearly as a function of $1/L^2$ for large $L$. The extrapolations 
of the DCA and finite size results meet as $L\rightarrow\infty$, approximating 
the value of the dressed Green function in the thermodynamic limit. 
The complementarity of DCA and finite size methods allows 
a determination of the thermodynamic limit of imaginary time Green
functions with unprecedented accuracy.

\section{Finite size versus the DCA FLEX results for the two dimensional 
Hubbard Model at half-filling}
The Hubbard model at half filling undergoes a phase transition to 
anti-ferromagnetic order at low temperatures. According to the 
Mermin-Wagner-Hohenberg theorem, for dimension $D=2$ the 
critical temperature is zero. However, as we continue to 
lower the temperature, close enough to zero, a pseudogap will appear
in the density of states (DOS) as a precursor 
to the anti-ferromagnetic phase (which has a full gap as its signature). 
An approach towards non-Fermi-liquid behavior is also visible in 
both the real and imaginary parts of the retarded self-energy. \cite{Deisz}

\begin{figure}
\epsfxsize=3.1in
\epsffile{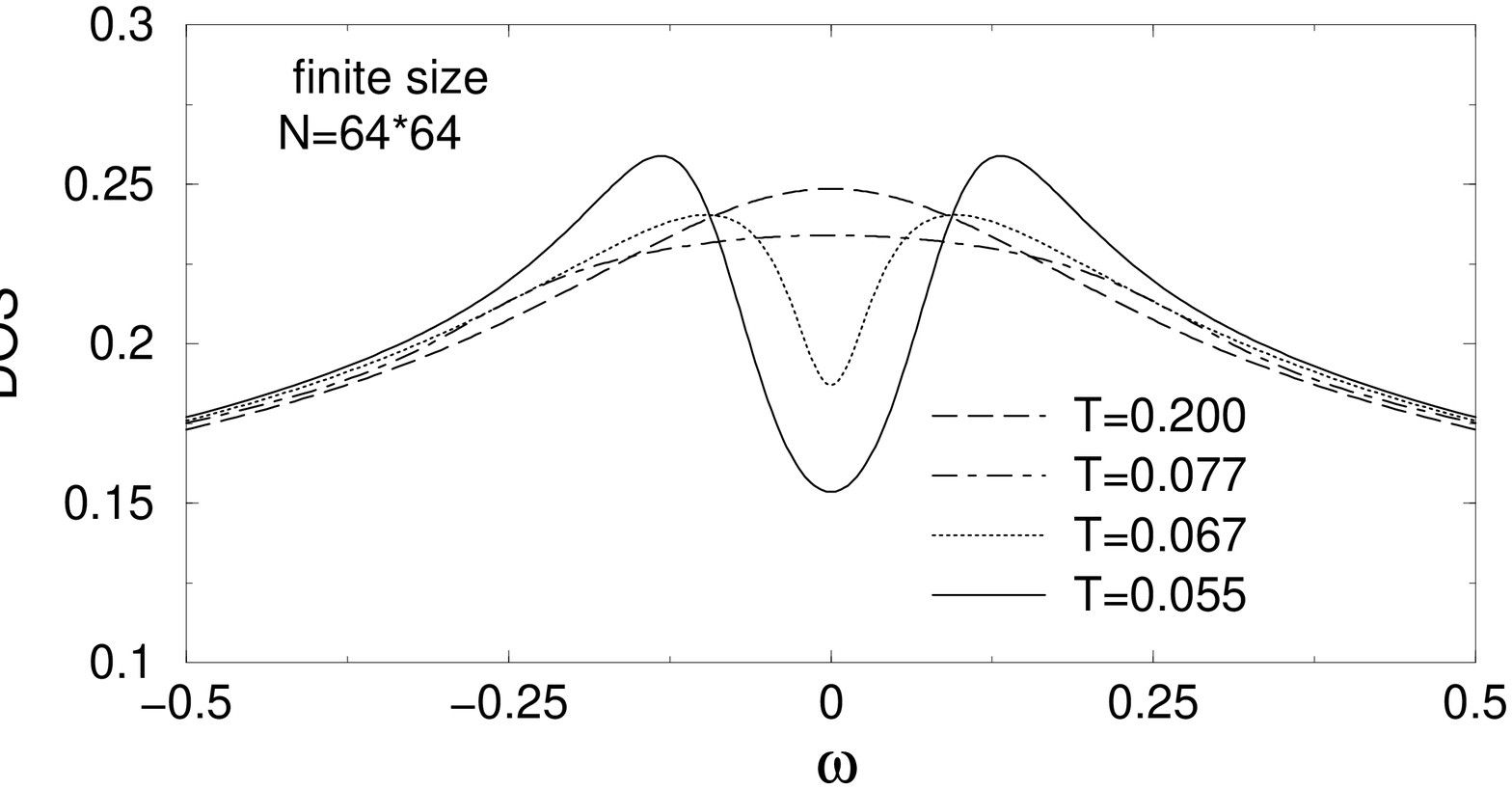}
\caption[a]{ The FLEX density of states (DOS) for a $32\times32$ 
finite size lattice with periodic boundary conditions (top) versus 
energy for $U/t=1.57$ at various temperatures, and for a $64\times64$ 
finite size lattice (bottom). The evolution of a week pseudogap is observed
at $T=0.067$ for the $64\times64$ lattice higher than $T=0.055$ 
for the $32\times32$ one and the pseudogap is also much broader 
for the $32\times32$ lattice ($128\times128$ finite size lattice 
results were obtained by Deisz {\it et al.} \cite{Deisz}).}
\label{DOS_finite}
\end{figure}

\begin{figure}
\epsfxsize=3.1in
\epsffile{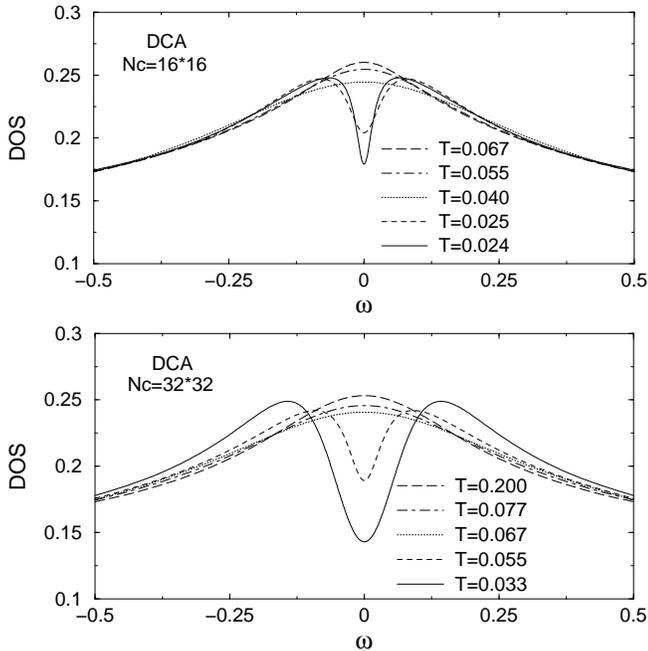}
\caption[a]{ The FLEX density of states (DOS) for a $16\times16$ 
DCA cluster (top) versus energy for $U/t=1.57$ at various temperatures,
and for a $32\times32$ DCA cluster (bottom). 
For the $32\times32$ cluster, the evolution of a weak pseudogap 
is observed at $T=0.055$ higher than $T=0.025$ for the $16\times16$ 
one and the pseudogap is also broader for the $32\times32$ cluster.}
\label{DOS_DCA}
\end{figure}

In Fig.~\ref{DOS_finite} and Fig.~\ref{DOS_DCA}, the densities of 
states (DOS) for lattices with finite sizes of 32$\times$32, 64$\times$64 and periodic 
boundary conditions and coarse-graining 
cluster sizes of 16$\times$16 and 32$\times$32 are plotted. We analytically 
continue the Green function in order to calculate the spectral function 
$A(k,\omega)$ and DOS using the Pad\'e approximation. \cite{Vidberg} 
In this approximation, we generate a continued fraction interpolating 
all the data points and use it as an analytic function of the 
Matsubara frequencies $\omega_n$. The analytic continuation is 
accomplished by substituting $\omega_n$ with $\omega + i\eta$ 
where $\eta$ is a small positive shift. However, the errors inherent 
in the numerical Fourier transform (FFT) and also the sharp 
high-frequency behavior of the Green function, FLEX potentials 
and the self-energy, limit the accuracy of the Pad\'e. The 
high-frequency behavior is improved by implementing a more 
authentic cut off scheme introduced by Deisz {\it et al.}\cite{DHS} 
in which the high-frequency tails of these quantities are
Fourier transformed analytically prior to any numerical FFT and 
added back to the FFT outputs afterwards. In addition to this 
high-frequency cut off improvement, we also check for the 
analyticity of the Pad\'e results in the upper-half frequency 
plane as a requirement for retarded physical quantities. This 
task is carried out by converting the continued fraction in the 
Pad\'e into a ratio of two polynomials. The complex roots of 
these two polynomials are obtained via the Jenkins-Traub root 
finder routine. \cite{Jenkins} Those orders of the Pad\'e for 
which there exist poles in the upper-half plane are omitted 
unless these poles are canceled by the roots of the numerator. 
The acceptable Pad\'e results correspond to the highest order 
with no uncompensated poles in the upper-half plane.

As seen in Fig.~\ref{DOS_finite}, by increasing the size of a 
finite lattice, the pseudogap occurs at higher temperatures 
and it also becomes less pronounced (sharper) as we approach 
the actual size of an infinite real lattice. The DCA yields a 
complementary behavior as shown in Fig.~\ref{DOS_DCA}. 
By increasing the coarse-graining cluster size, similar to 
the finite size lattices, the pseudogap is shifted towards 
higher temperatures. However, unlike the finite size lattices, 
for the DCA the precursor becomes more pronounced (broader) 
as the cluster increases in size because the size of the lattice 
remains constant and the correlations are limited to the 
cluster size.
Thus comparatively, {\em{the DCA underestimates the gap while 
the finite size calculation overestimates it}}.

By comparing the results in Fig.~\ref{DOS_finite} and
Fig.~\ref{DOS_DCA} 
one may see that the $32\times32$ DCA cluster yields more realistic 
physics than the corresponding $32\times32$ finite size lattice. The 
$64\times64$ finite size lattice results are also close to those for 
the $32\times32$ DCA cluster at slightly lower $T$ (eg, $T=0.055$ for 
the finite size and $T=0.033$ for the DCA). However since the 
sizes of clusters are considerably smaller than the sizes of lattices, 
the DCA significantly reduces the complexity of the problem and 
consequently the CPU time. In terms of the CPU time, the FLEX with 
the numerical Fourier transforms scales as ${\cal N}ln{\cal N}$ 
where ${\cal N}$ is the product of the total number of 
Matsubara frequencies and the $k$ points in the first Brillouin zone. 
Hence, using a $32\times32$ cluster in place of a $64\times64$ lattice 
both having $1024$ Matsubara frequency points roughly reduces the 
CPU time by a factor of $4.4$. If the DCA cluster size $N_c$ equals 
the size of the finite lattice $N$, the DCA requires somewhat more CPU 
time than the finite size lattice due to course-graining. Nevertheless, 
comparatively, for large finite lattices such as $64\times64$, 
the lattice size contributions to the CPU time significantly 
dominate the coarse-graining ones in a $32\times32$ cluster. 
Thus, the $32\times32$ DCA cluster is much faster.   
\begin{figure} 
\epsfxsize=3.1in
\epsffile{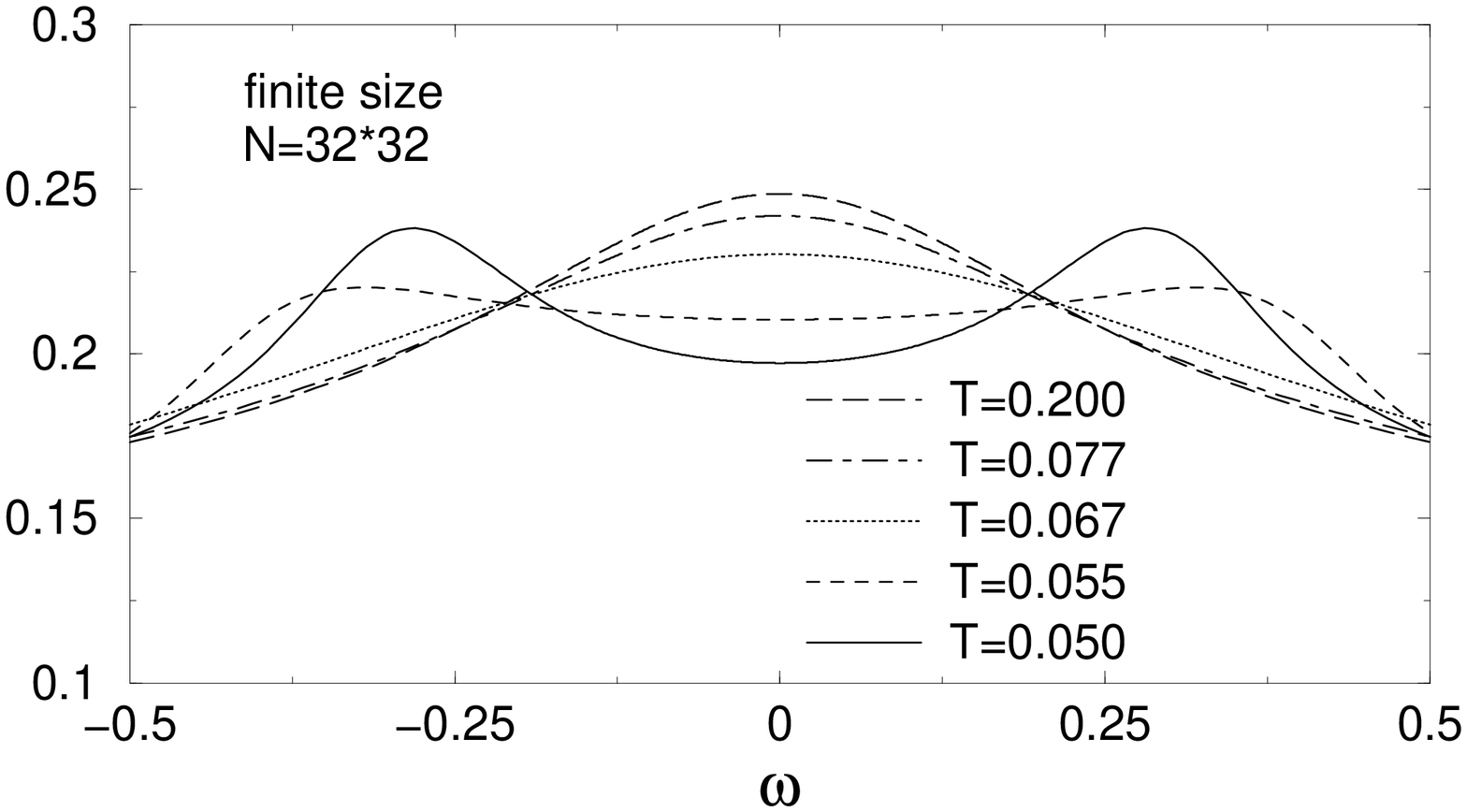}
\end{figure}
\vspace{-1.0cm}

\begin{figure}
\epsfxsize=3.0in
\epsffile{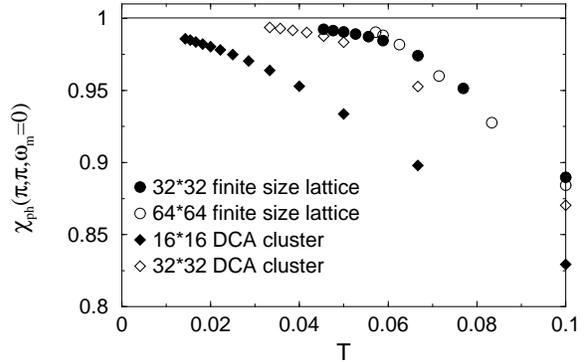}\hspace{-1.0cm}
\vspace{0.1cm}
\caption[a]{ The FLEX particle-hole bubble ($\chi_{ph}$) at 
$Q = (\pi,\pi)$ and $\omega_m = 0$ for a $16\times16$ and 
$32\times32$ DCA clusters (diamonds) and also for $32\times32$ 
and $64\times64$ finite size lattices with periodic boundary conditions (circles) versus temperature T 
at $U/t=1.57$. The relatively rapid saturation of the finite size 
lattice $\chi_{ph}$  compared to the more gradual approach of the 
DCA results towards unity is manifest ($128\times128$ finite size 
lattice results were obtained by Deisz {\it et al.} \cite{Deisz}).}
\label{chi-comp}
\end{figure}

\begin{figure}
\epsfxsize=3.0in
\epsffile{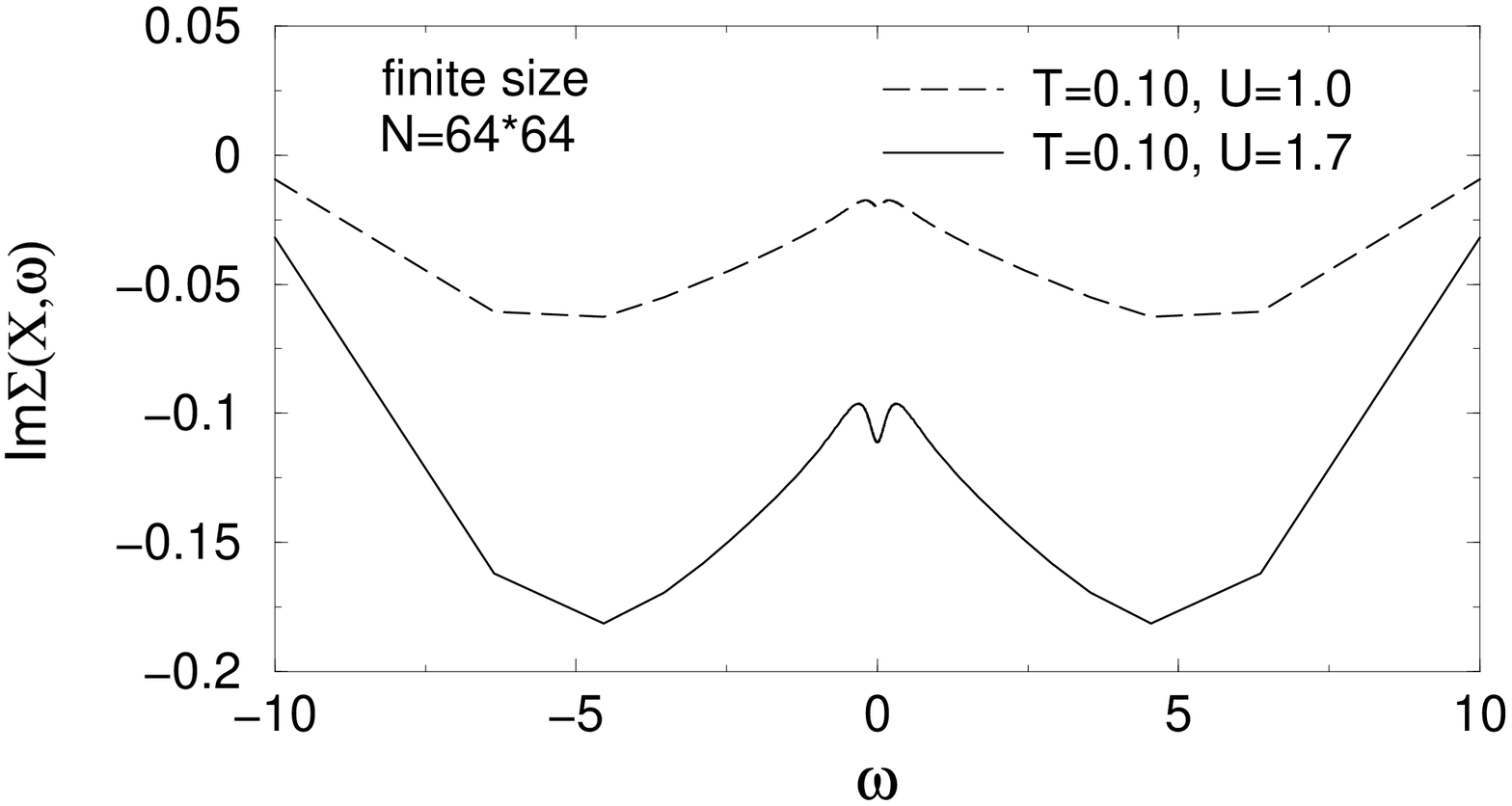}
\caption[a]{ The FLEX real (top two) and imaginary (bottom two) parts 
of the self-energy at the X point for $32\times32$ and $64\times64$ 
finite size lattices with periodic boundary conditions versus energy for 
$T=0.10$ and two different interactions $U$. The inverted peak at $\omega=0$ 
in the imaginary part and the positive slope in the real part are both signatures of 
non-Fermi-liquid behavior. For smaller lattice sizes these signatures 
are more pronounced ($128\times128$ finite size lattice results were 
obtained by Deisz {\it et al.} \cite{Deisz}).}
\label{Sig-Fin}
\end{figure}

\begin{figure}
\epsfxsize=3.0in
\epsffile{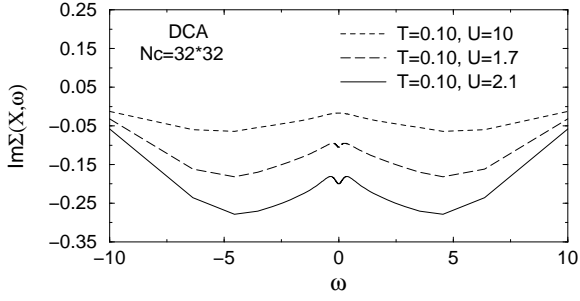}
\caption[a]{ The FLEX real (top two) and imaginary (bottom two) parts 
of the self-energy at the X point for $16\times16$ and $32\times32$ 
DCA clusters versus energy for $T=0.10$ and two different interactions $U$. 
The non-Fermi-liquid features occur at slightly higher interactions 
compared to the finite size lattices. Upon increasing the cluster 
size, these features become more pronounced (complementary to the 
finite size lattices with periodic boundary conditions).}
\label{Sig-DCA}
\end{figure}

The FLEX often has difficulty reaching low temperatures. This is due to 
the fact that the $\chi_{ph}$ defined in Eq.~\ref{chiph} approaches 
unity as the temperature drops which in turn causes the $V^{(ph)}$ 
in Eq.~\ref{eq:flex-pots1} to diverge. In the DCA, $\chi_{ph}$
approaches unity more slowly, allowing the calculations to reach 
lower temperatures. One has to note that in the FLEX, the 
$\chi_{ph}^{DCA}$ is defined as follows
\begin{eqnarray}
\chi_{ph}^{DCA}(\Q,\omega_{n})=\nonumber\\&&\hspace{-2.5cm}
-U(T/N_c)\sum_{\K}\sum_{m}\bar G(\K+\Q,\omega_{n}+\omega_{m})
\bar G(\K,\omega_{m})\,,
\label{chiph_DCA}
\end{eqnarray}
with $\bar G$ defined in Eq.~\ref{eq:cgGDCA}. Fig.~\ref{chi-comp} 
illustrates the saturation of $\chi_{ph}$ for both the DCA 
and finite size lattices. The $\chi_{ph}$ for the $32\times32$ 
finite size lattice (filled circles) saturates at higher temperatures 
compared to the $32\times32$ DCA cluster (open diamonds) indicating 
that in the DCA, for a certain cluster size, the precursor to the 
phase transition can evolve to lower temperatures compared to a 
finite size lattice of the same size with periodic boundary conditions. 
However, for the correlation length $\xi > L$ the DCA approximation breaks down 
and replacing the self-energy by its coarse-grained counterpart 
is no longer accurate. Here, the DCA takes on significant mean-field character.
\begin{figure} 
\epsfxsize=3.1in
\epsffile{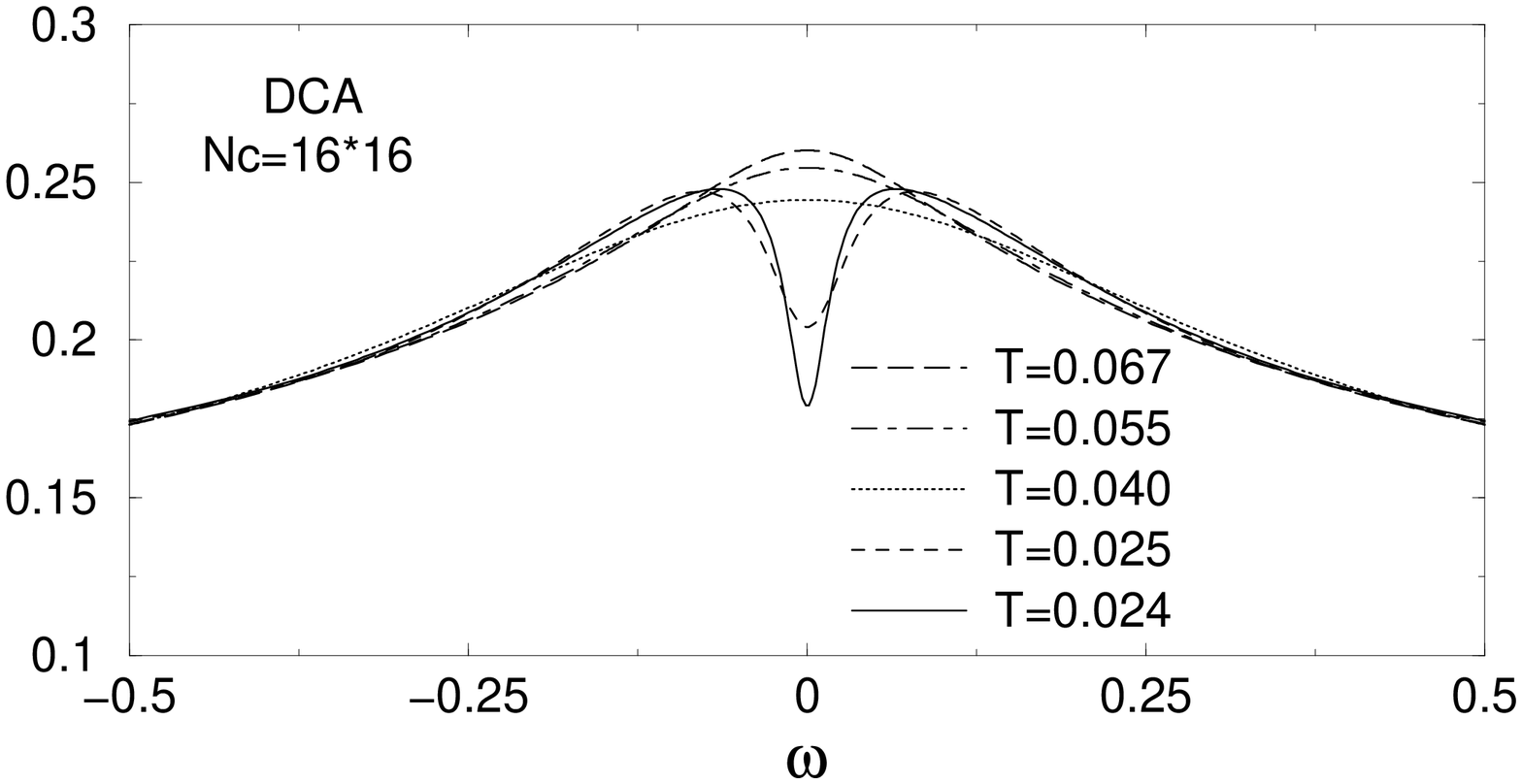}
\end{figure}
\vspace{-1.0cm}
\vspace{-0.10cm}
Another feature of the Hubbard model near half-filling verified by the 
FLEX \cite{Deisz} is non-Fermi-liquid behavior.  Here, this is studied 
by increasing the electron electron interaction U at a constant 
temperature. In Fig.~\ref{Sig-Fin} and Fig.~\ref{Sig-DCA} the real and 
imaginary parts of the self-energy at the X point (on the 
non-interacting Fermi surface) have been plotted versus energy 
for finite size lattices and the DCA respectively. 
As the interaction is increased, the negative slope in the 
real part turns positive around $\omega=0$ which is inconsistent 
with the requirement that the renormalization factor 
$[1- \partial Re\Sigma(\k_F,\omega)/\partial\omega|_{\omega=0}]^{-1}$ 
should be smaller than unity in the Fermi-liquid theory. 
There also appears an anomalous inverted peak in the 
imaginary part at $\omega=0$.
 
As presented in Fig.~\ref{Sig-Fin}, by increasing the length of the 
finite size lattice, the sharpness of the non-Fermi-liquid features is 
reduced. The same features for the DCA in Fig.~\ref{Sig-DCA} are slightly 
less pronounced and in a complementary fashion to the finite size 
lattices, their sharpness is enhanced by increasing the size of the
cluster. Thus, again the DCA underestimates the non-Fermi-liquid 
features while the finite size calculation overestimates it.
\vspace{0.1cm}

\begin{figure}
\epsfxsize=3.0in
\hspace{0.4cm}\epsffile{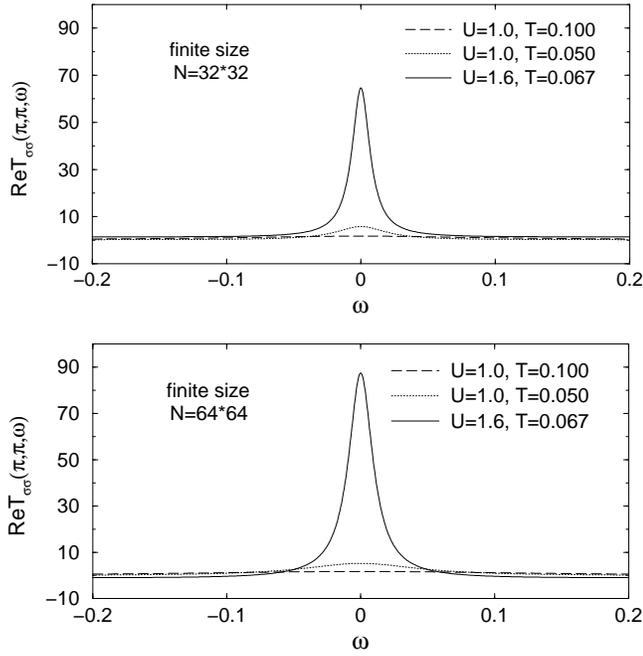}
\caption[a]{ The FLEX real part of the T-matrix at $Q=(\pi,\pi)$ 
versus energy for $32\times32$ (top) and $64\times64$ (bottom) finite 
size lattices with periodic boundary conditions. For the $32\times32$ finite size lattice at $T=0.067$ and $U/t=1.6$ (solid), we observe a huge peak due to 
the formation of a pseudogap in the DOS. By increasing the lattice size to 
$64\times64$ a higher and broader peak occurs at the same temperature.}
\label{T.Matrix-Fin}
\end{figure}

\begin{figure}
\epsfxsize=3.0in
\hspace{0.4cm}\epsffile{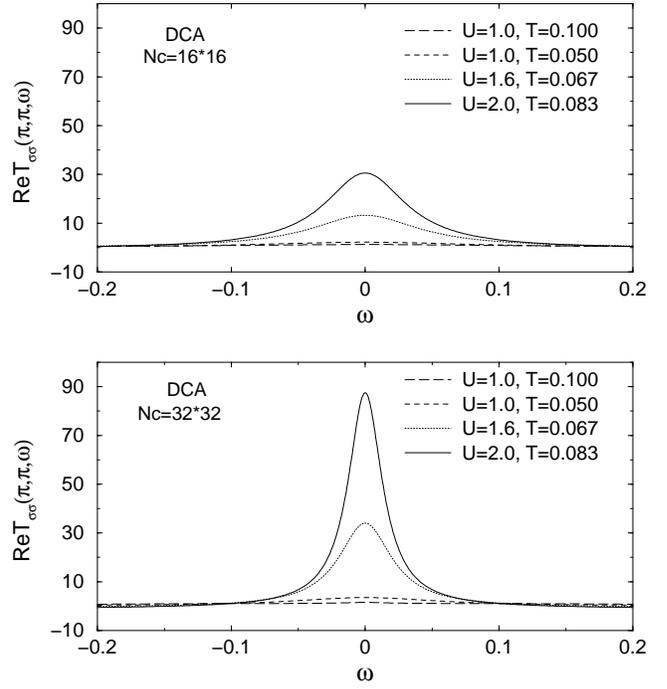}
\caption[a]{ The FLEX real part of the T-matrix at $Q=(\pi,\pi)$ 
versus energy for $16\times16$ (top) and $32\times32$ (bottom) DCA 
clusters. Peaks similar to the finite size lattices appear at 
slightly higher interactions and lower temperatures. By increasing 
the cluster size, the peaks grow higher (similar to finite size 
lattices) and also sharpen (complementary to the finite size lattices).}
\label{T.Matrix-DCA}
\end{figure}

Earlier in Fig.~\ref{chi-comp} it was shown that the FLEX 
particle-hole bubble $\chi_{ph}(q,\omega_m=0)$ at $q=Q=(\pi,\pi)$ 
approaches unity as the temperature is lowered. This causes the 
spin-fluctuation $T$ matrix  
\begin{eqnarray}
\label{eq:T-matrix}
T_{\sigma,\sigma}(q,\omega_{m}) = 
\frac{3}{2}\big[\frac{\chi_{ph}(q,\omega_{m})^2}{1 - 
\chi_{ph}(q,\omega_{m})}\big]\,,
\end{eqnarray}
which is just the third term in $V^{(ph)}(q,\omega_{m})$ in 
Eq.~\ref{eq:flex-pots1} to peak around the $(Q,\omega_m=0)$ point. 
For real frequencies, $T_{\sigma,\sigma}(Q,\omega)$ has a sharp peak 
around $\omega=0$. Since $T_{\sigma,\sigma}(Q,\omega)$ is only used 
to construct the irreducible self-energy, within the DCA it is 
constructed from coarse-grained Green functions. Thus, the DCA 
counterpart of Eq.~\ref{eq:T-matrix} is obtained by only replacing 
the $\chi_{ph}$ with $\chi_{ph}^{DCA}$ defined in
Eq.~\ref{chiph_DCA}. 
Fig.~\ref{T.Matrix-Fin} and Fig.~\ref{T.Matrix-DCA} show how this 
peak sharpens as the temperature decreases or interaction increases 
for finite size lattices and the DCA respectively. In the $64\times64$ 
finite size lattice (c.f. Fig~\ref{T.Matrix-Fin} bottom), the peak 
continues to develop as the temperature is lowered and the 
interaction is raised. At $T=0.067$ and $U=1.6$ where there exists 
a pseudogap in the DOS, the peak undergoes a significant growth 
compared to the other graphs shown in the same figure. The
$32\times32$ (c.f. Fig~\ref{T.Matrix-Fin}.top) lattice presents the 
same behavior with a slightly sharper but shorter peak.

The DCA also illustrates the same type of peaks at a slightly higher 
interaction and lower temperatures for the $16\times16$ cluster 
(c.f. Fig~\ref{T.Matrix-DCA} top). Increasing the cluster size to 
$32\times32$ gives rise to higher peaks similar to the finite size 
lattices case but unlike the finite size lattices peaks become sharper 
as the size is increased.  

\begin{figure}
\epsfxsize=3.0in
\epsffile{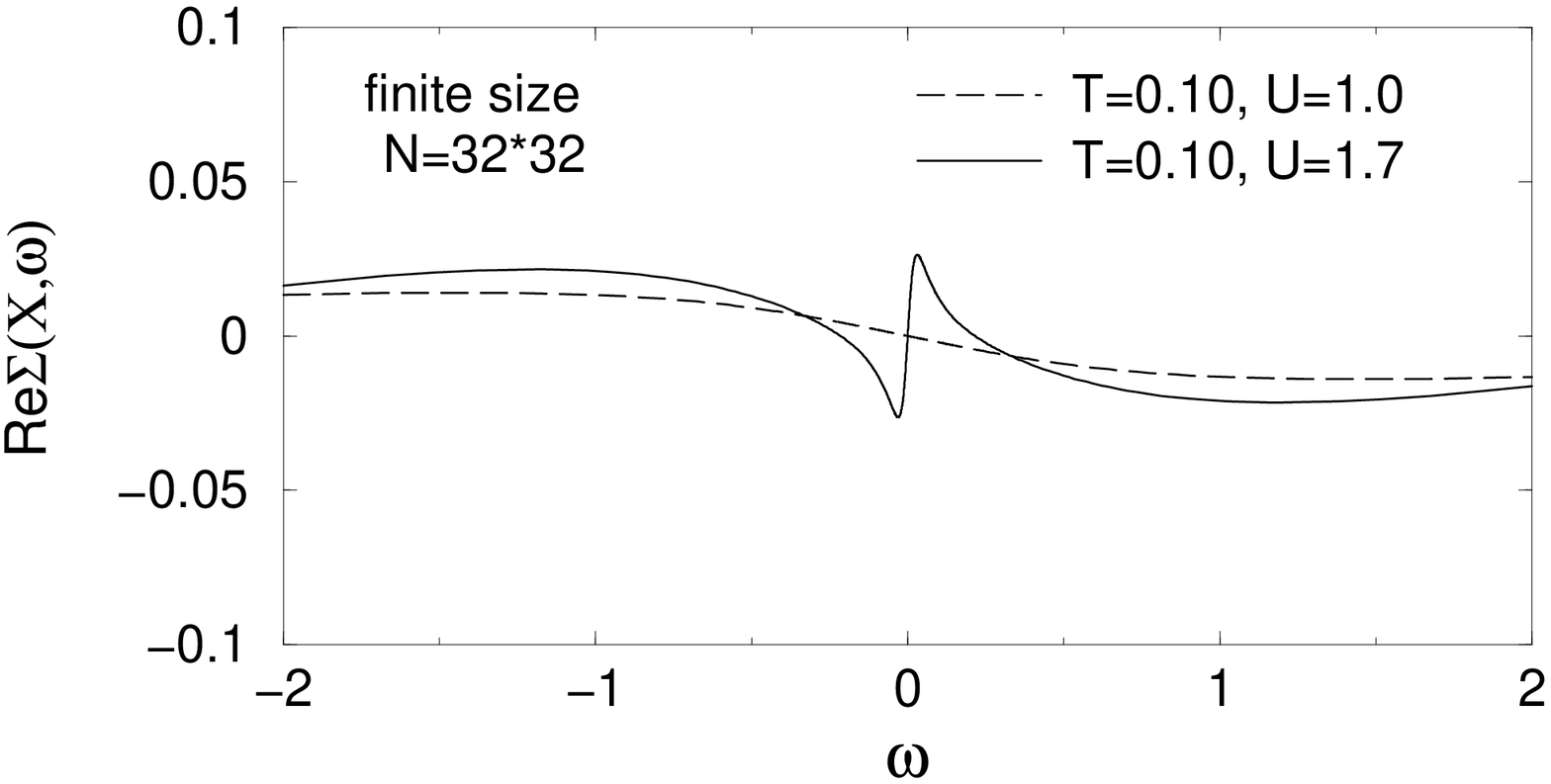}
\end{figure} 
\vspace{-1.1cm}
\begin{figure}
\epsfxsize=3.0in
\epsffile{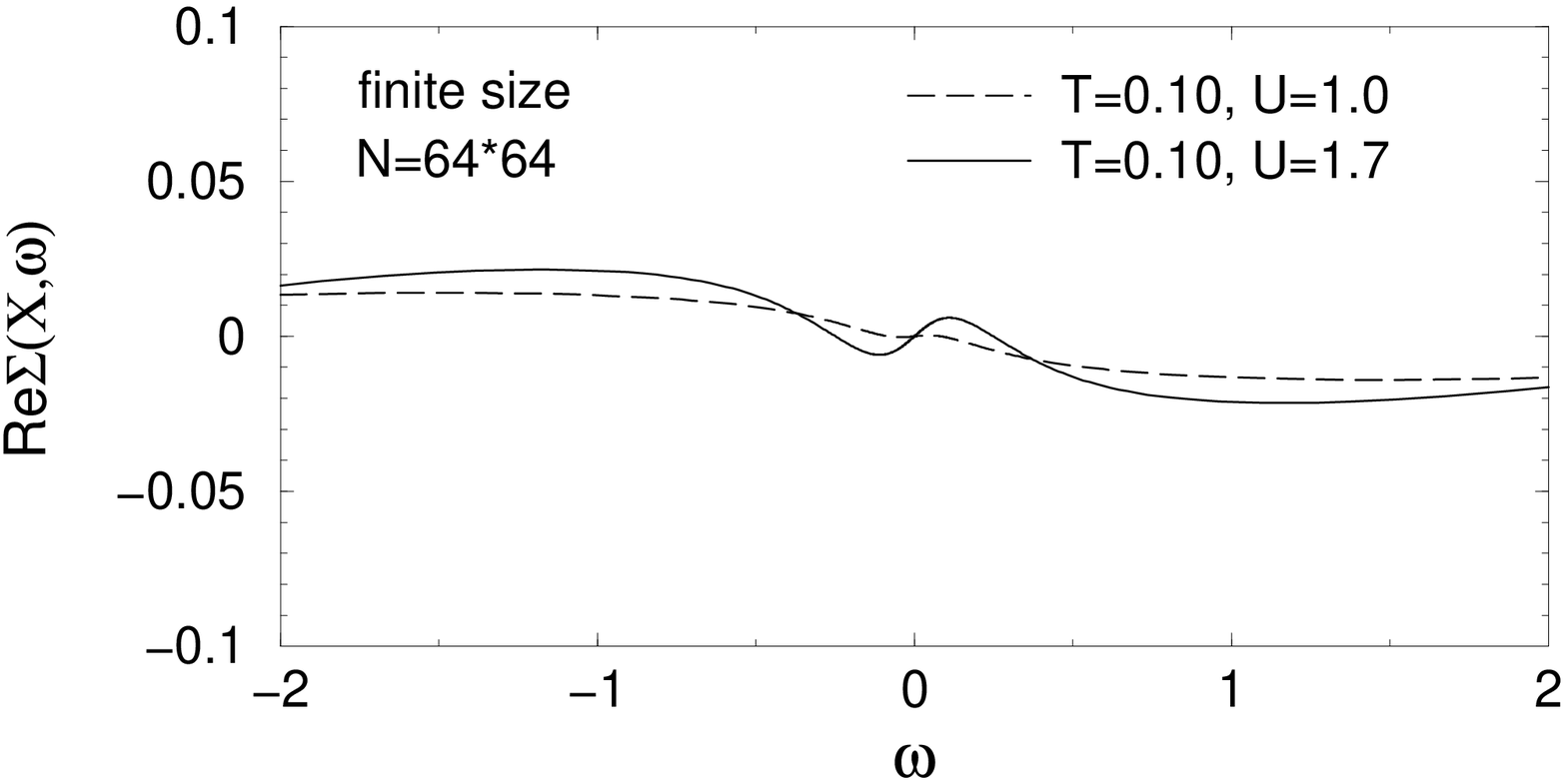}
\end{figure} 
\vspace{-1.1cm}
\begin{figure}
\epsfxsize=3.0in  
\epsffile{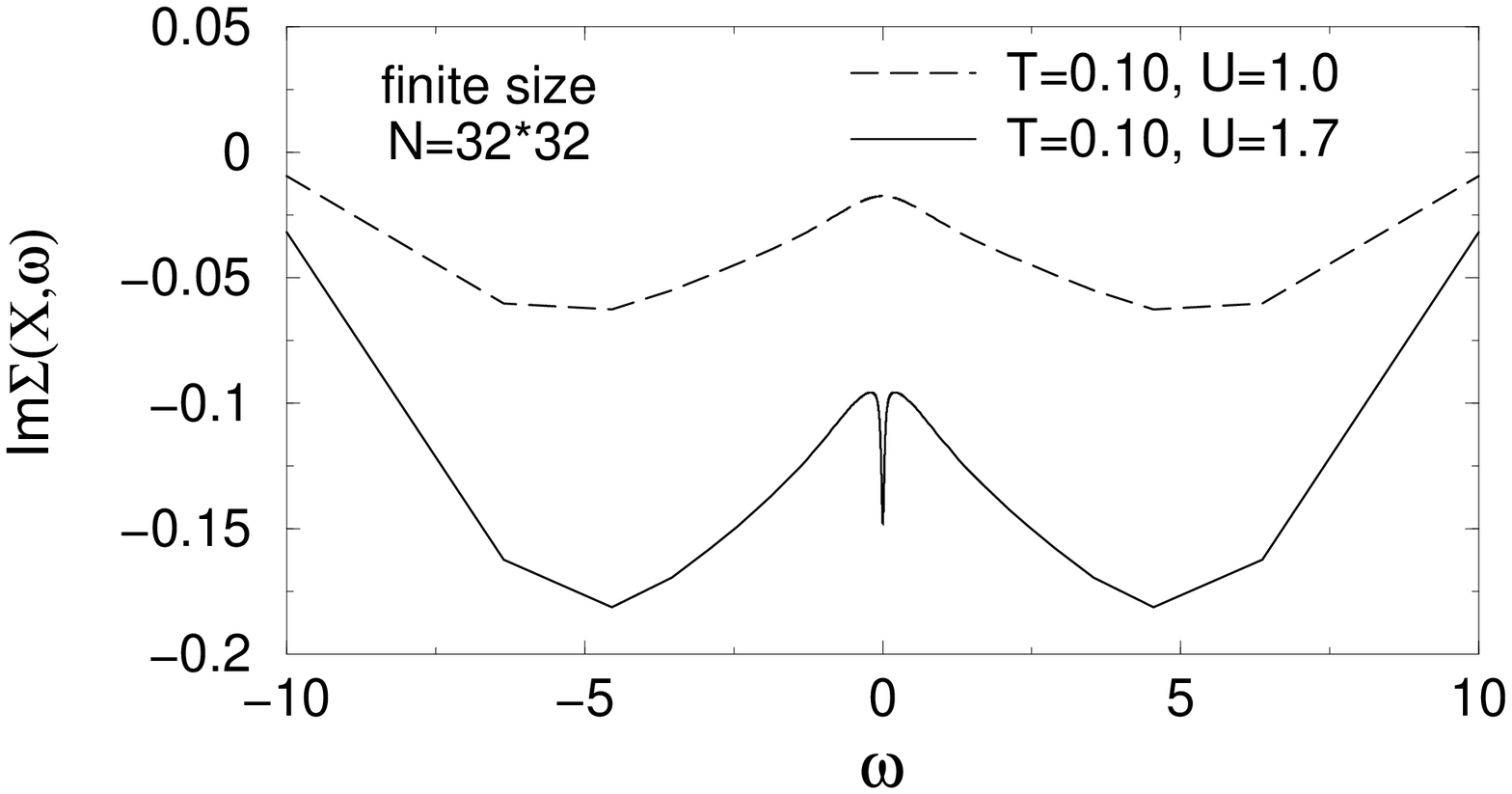}
\end{figure} 
\vspace{-1.1cm}
\vspace{-0.6cm}
\begin{figure}
\epsfxsize=3.0in
\epsffile{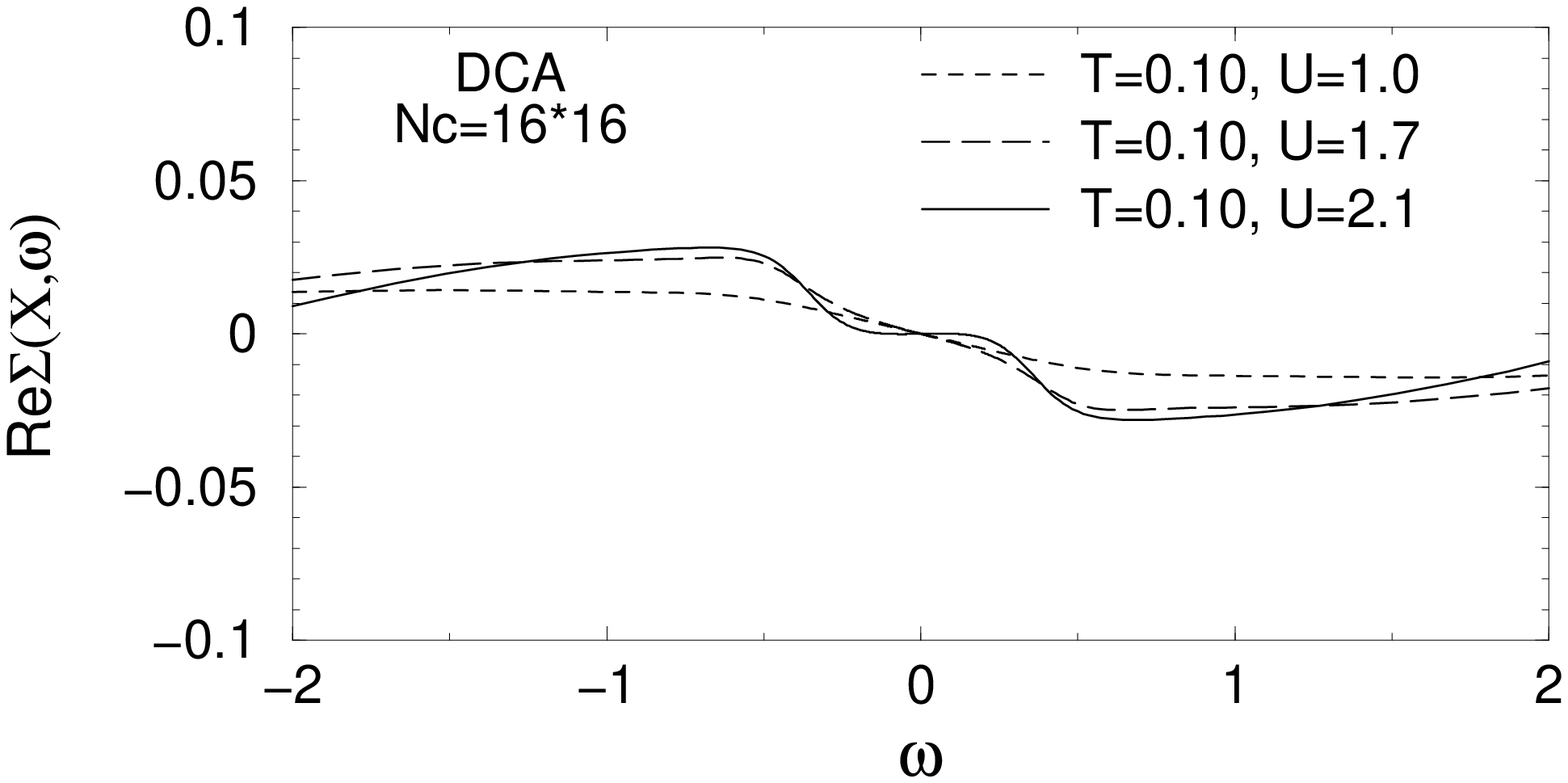}
\end{figure}
\vspace{-1.0cm}
\begin{figure}
\epsfxsize=3.0in
\epsffile{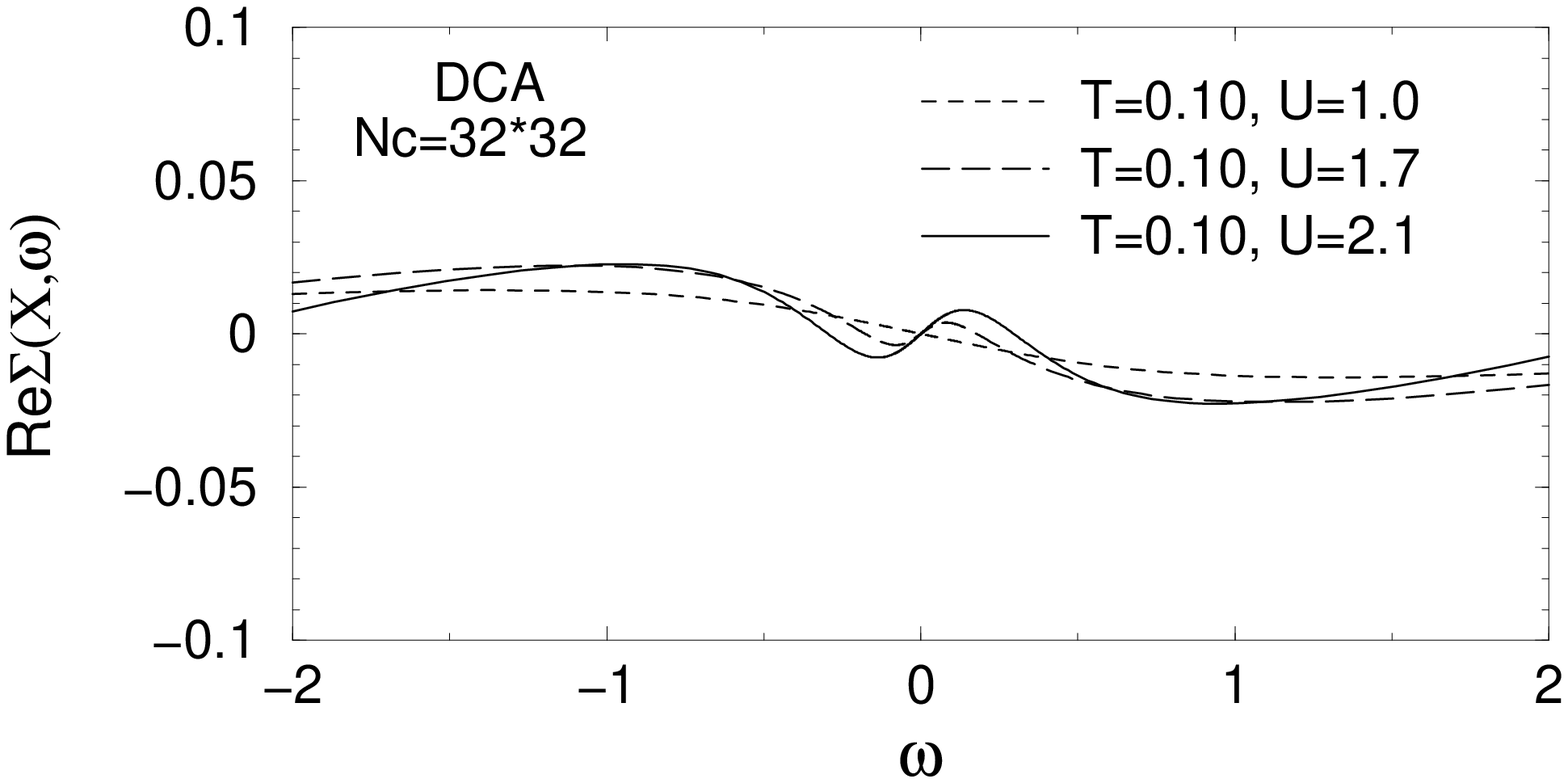 }
\end{figure}
\vspace{-1.0cm}
\begin{figure}
\epsfxsize=3.0in
\epsffile{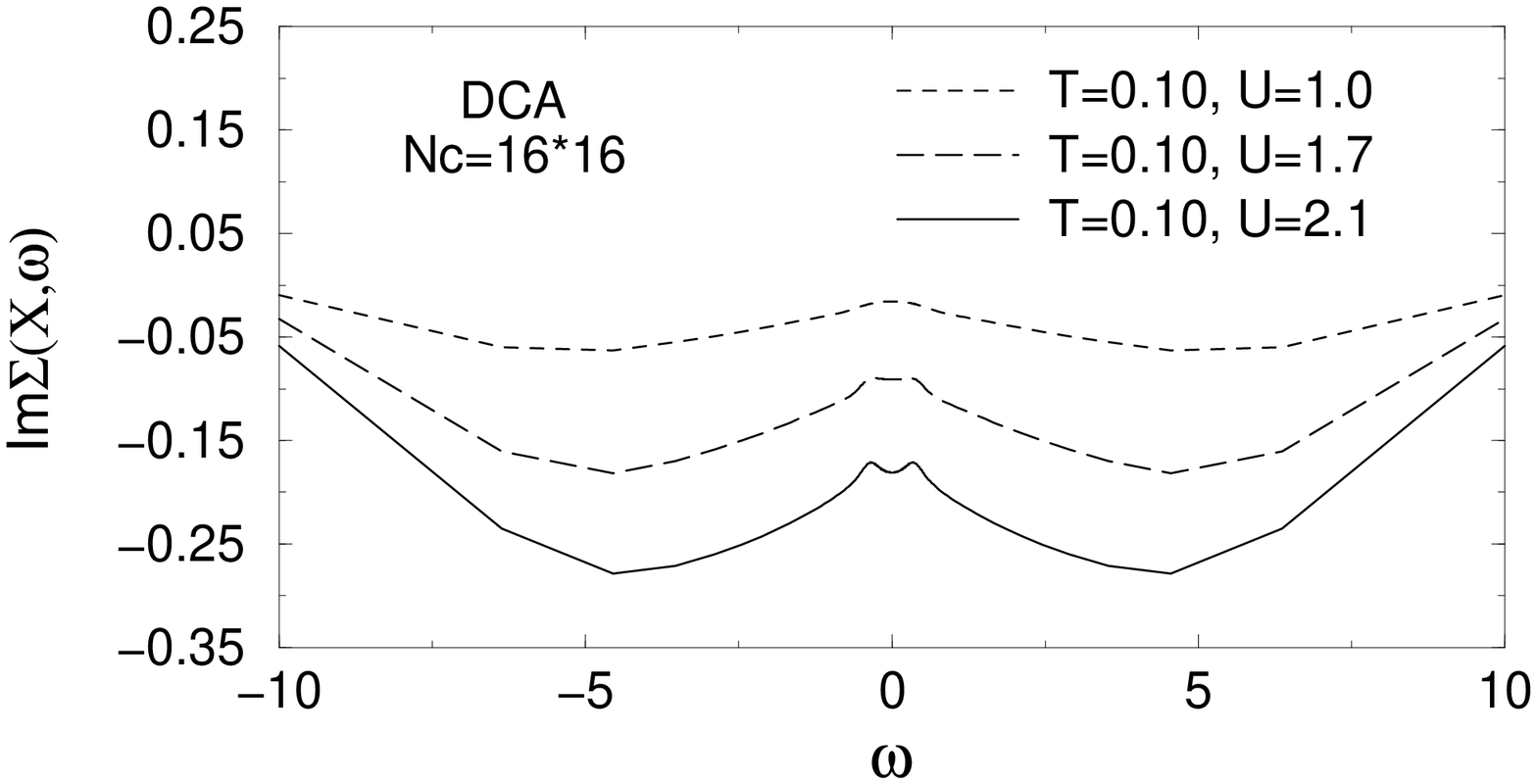}
\end{figure}
\vspace{-1.0cm}
All the results illustrated in this section indicate the complementarity 
of the DCA to the finite size lattice scheme. It is also observed that 
the DCA is capable of reproducing relatively the same physics as the 
finite size FLEX at slightly different parameters but a lower 
CPU cost. The combination of these two facts makes this technique a 
good candidate to be employed in the numerical treatment of a wide range 
of many-body problems.  
\begin{figure}
\epsfxsize=3.0in
\hspace{0.4cm}\epsffile{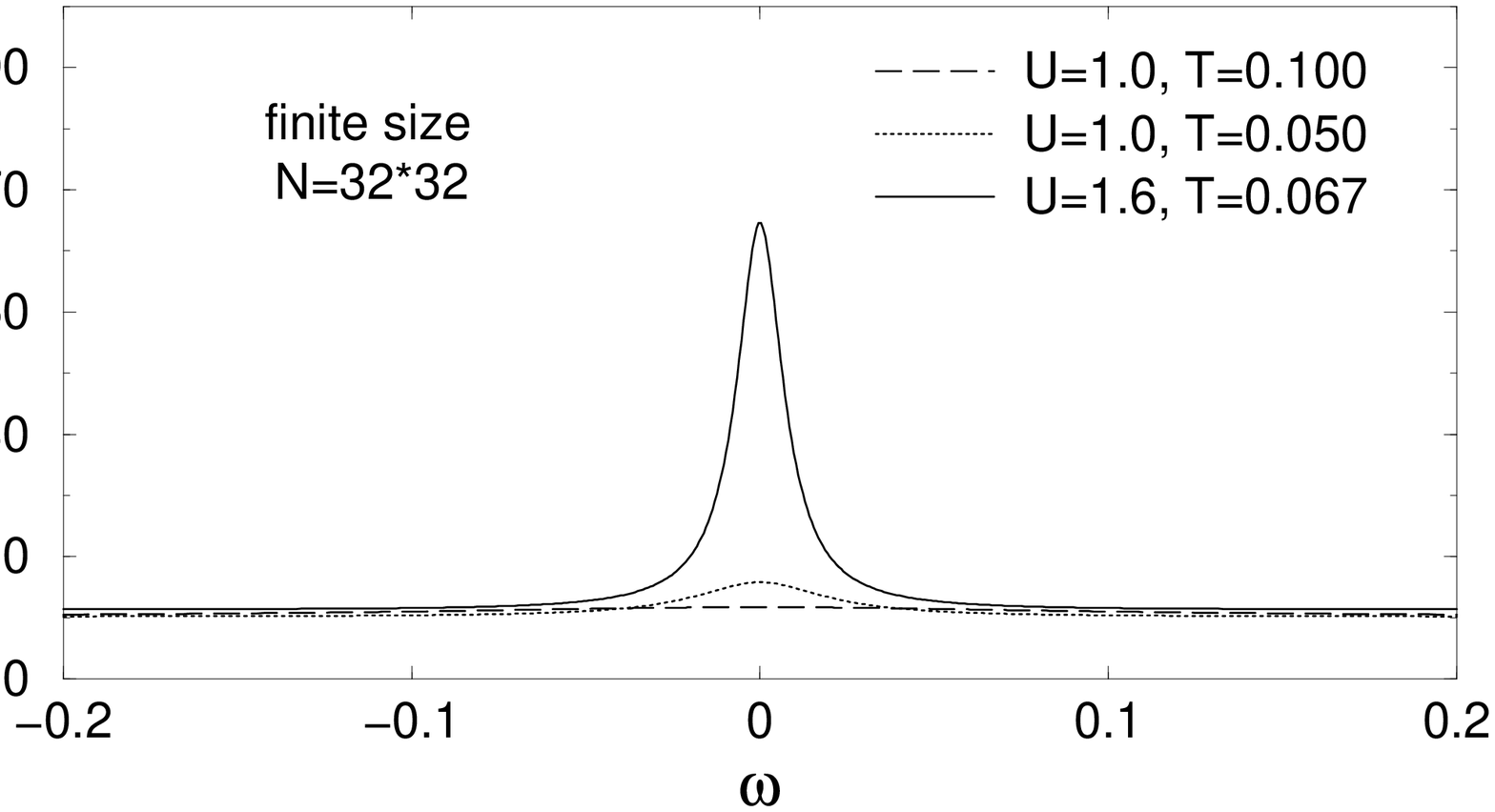}
\end{figure}
\vspace{-0.7cm}
\begin{figure}
\epsfxsize=3.0in
\hspace{0.4cm}\epsffile{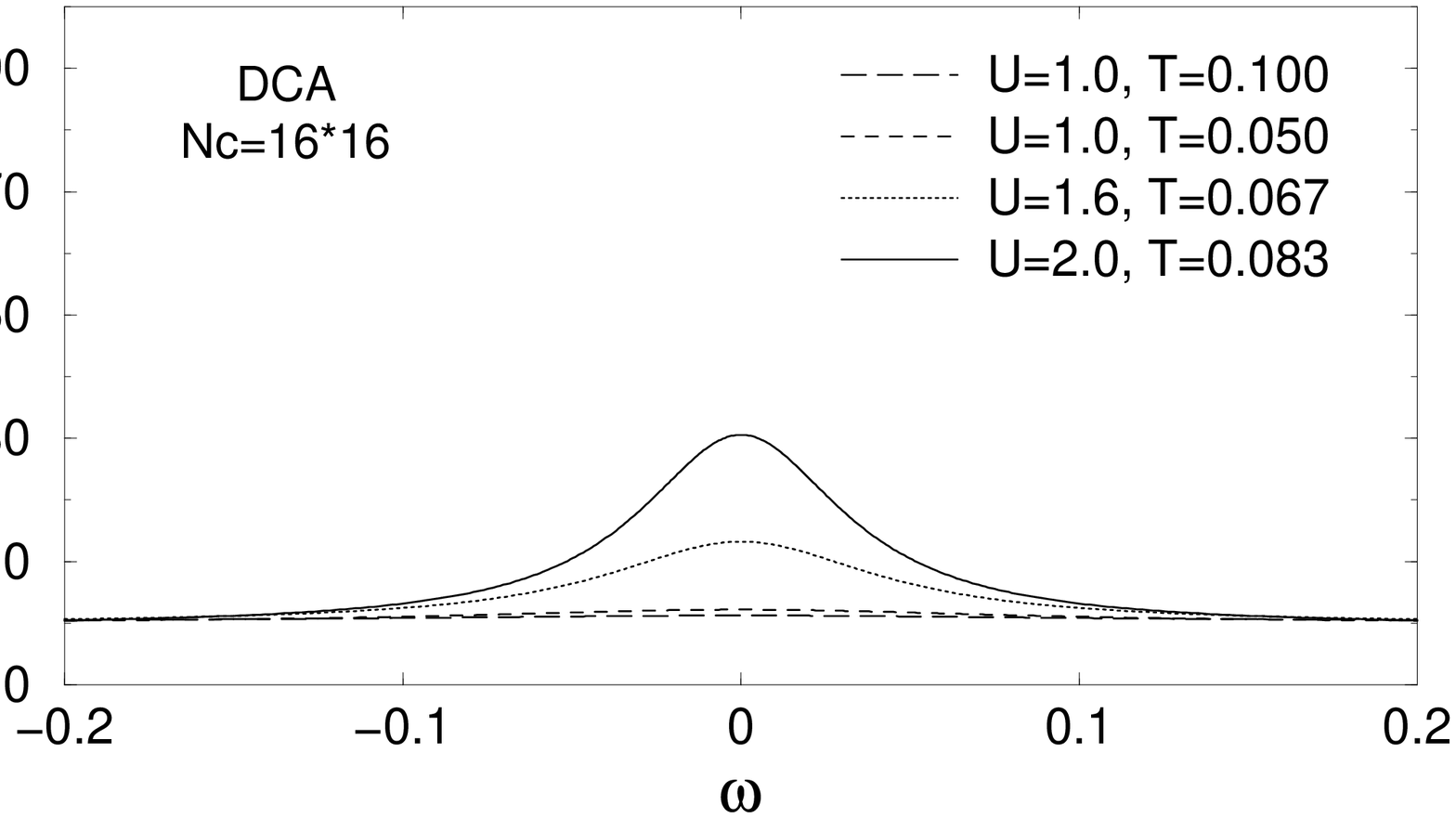}
\end{figure}
\vspace{-0.7cm}
  
\begin{figure}
\epsfxsize=3.0in
\hspace{-0.2cm}\epsffile{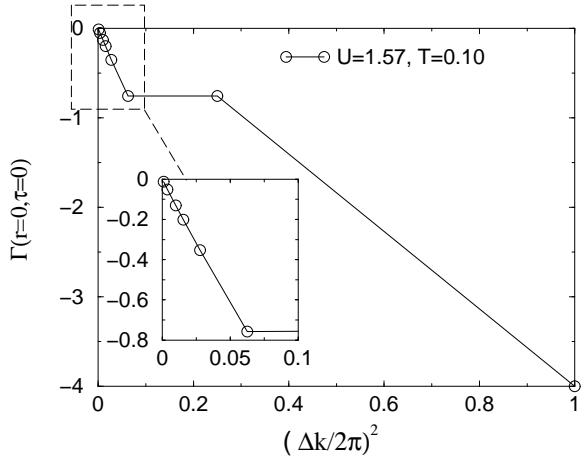} 
\vspace{0.1cm}
\caption[a]{The host function $\Gamma(r=0,\tau=0)$ versus 
$(\Delta  k)^2$. The linear 
behavior beyond $N_c=4$ is manifest, see also the inset. 
}
\label{Gamma-DCA}
\end{figure}

\begin{figure}
\epsfxsize=1.3in
\hspace{1.0in}\epsffile{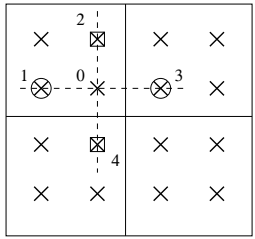} 
\vspace{0.3cm}
\caption[a]{DCA clusters in the real space lattice with $N_c=4$. 
Due to the periodicity of the clusters, points 2 and 4 in squares 
and 1 and 3 in circles are equivalent and therefore point 0 observes 
its nearest neighbors twice in each direction. 
}
\label{NC2}
\end{figure}

Lastly, to better understand the effect of the DCA cluster embedded in 
a Fermionic bath, we can rewrite the course grained Green function defined 
in  Eq.~\ref{eq:cgGDCA} as
\vspace{0.1cm}
\begin{equation}
\label{eq:Gamma}
\bG(\K,z)=\frac{1}{z-{\bar\epsilon}_{\K}-\Sigma(\K,z)-\Gamma(\K,z)}\,.
\end{equation} 
\vspace{0.1cm}
\hspace{-0.1cm}where 
${\bar\epsilon}_{K}=N_{c}/N\sum_{\tk}{\epsilon}_{\K+\tk}$ and 
$\Gamma(\K,z)$ is the host function. 
Maier {\it et al.} \cite{DCA_maier1}, define 
\begin{equation}
\label{eq:Tfunc}
t_{\K+\tk}={\epsilon}_{\K+\tk}-{\bar\epsilon}_{\K}\,,
\end{equation} 
whereby $\Gamma(\K,z)$ can be expressed as 
\vspace{0.1cm}
\begin{equation}
\label{eq:Gamma-T}
\Gamma(\K,z)=\frac{\frac{N_c}{N}\sum_{\tk}{t^2}_{\K+\tk}
G(\K+\tk,z)}{1+\frac{N_c}{N}\sum_{\tk}{t}_{\K+\tk}G(\K+\tk,z) }\,.
\end{equation} 
\vspace{0.1cm}
By Taylor expanding $t_{\K+\tk}$ around the cluster points $\K$ it 
is found that $t_{\K+\tk}\sim{\cal O}(\Delta k)$ with $\Delta k =2\pi/L$.
Thus, Eq.~\ref{eq:Gamma-T} yields $\Gamma(\K)\sim{\cal O}((\Delta k)^2)$
as $\Delta k \to 0$.  To illustrate this, we calculate 
$\Gamma(r=0,\tau=0)$ by summing over all the $\K$ points and  
$\omega_n$ frequencies and plot it versus 
$(\Delta k)^2$. Fig.~\ref{Gamma-DCA} illustrates this linear behavior 
for $N_c\geq 16$. $N_c=1$ holds complete mean field characters and no 
non-local fluctuations. $N_c=4$ is  anomalous as explained in 
an article by Betts {\it et al.} \cite{Betts} There, the finite 
size cubic lattices with less than six (four in $2D$ square lattices) 
distinct nearest neighbors per each site are not used in finite size 
scalings for estimating the physical properties of models like the 
spin one half XY ferromagnet or the Heisenberg antiferromagnet. 
For $N_c=4$, because of the periodicity of the clusters, each 
cluster point is surrounded by two identical nearest neighbors in 
every direction and therefore has only two distinct nearest neighbors 
(c.f.~Fig.~\ref{NC2}).
Thus, the effect of fluctuations are overestimated.
\cite{Th-Maier} For $N_c > 4$ there is no such anomaly and hence, 
all the points present the linear behavior proven above.  
Nevertheless, calculations with $N_{c}=4$ do a reasonable job in 
capturing the qualitative effects of corrections to the DMFA. 
\vspace{-0.4cm}
\section{Microscopic theory of the DCA}
\label{microscopic}
\begin{figure}
\epsfxsize=3.5in
\epsffile{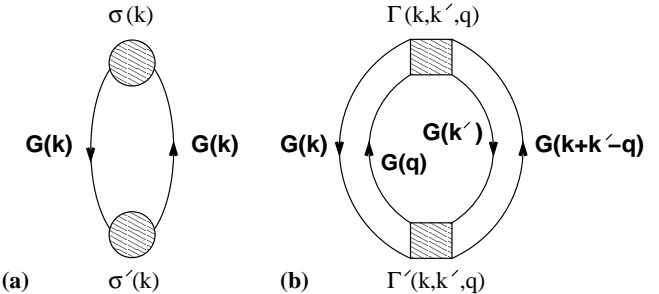}
\caption[a]{\em{(a) typical non-compact (non-skeletal) and (b) typical 
compact (skeletal) diagrams.}}
\label{comp_noncomp}
\end{figure}

In section~\ref{Fluct-App} we defined the thermodynamic potential 
functional difference $\Delta\Omega(T,\mu)$ in terms of the Green 
function, self-energy, and $\Phi[G]$ (c.f. Eq.~\ref{eq:thermPot}). 
In Eq.~\ref{eq:FLEX_phi}, $\Phi[G]$ includes all the compact
(skeletal) 
Feynman diagrams and the rest incorporates the entire non-compact 
contribution. \cite{agd} Typical compact and non-compact diagrams 
are illustrated in Fig.~\ref{comp_noncomp}. The non-compact diagram 
(a) consists of two self-energy pieces $\sigma$ and $\sigma'$
connected 
with two one-particle Green functions. Removing these two Green 
functions would split the diagram into two separate pieces. 
In the compact diagram (b) two vertex parts $\Gamma$ and $\Gamma'$
with four Green functions are connected together. One can not split 
this type of diagrams into two separate pieces by just removing 
two one-particle Green functions. As mentioned earlier, 
in the DCA, we employ coarse-grained Green functions to construct 
only the compact diagrams. The Green functions in non-compact 
diagrams are calculated directly using Eq.~\ref{G_DCA} in which 
the self-energy $\Sigma_{DCA}(\K,z)$ is coarse-grained 
(the circles at the top and the bottom of the non-compact 
diagram in Fig.~\ref{comp_noncomp}.~(a))
\vspace{-0.3cm}
Earlier in a shorter article for the Hubbard model in particular 
\cite{Aryanpour}, we showed both analytically and numerically that 
the error produced by coarse-graining the non-compact diagrams is 
significantly larger than the error produced by coarse-graining the 
compact ones. Here, we would like to give a more general argument 
in real space. We wish to emphasize two points in this new 
approach. First, since the derivation of the DCA in this section relies 
only upon the exponential fall off of the Green function as a function of 
distance, it is far more intuitive than the momentum space argument 
in Ref.~\cite{Aryanpour}.  Second, it ties the derivation of the 
DCA to the original derivation of the DMFA in the limit of infinite
dimensions, where similar arguments are employed.\cite{Metzner}

The exponential fall off behavior occurs naturally in high dimensions.
In the tight-binding Hamiltonian (c.f. Eq.~\ref{Hubbard2}), 
the factor $t$ corresponds to the hopping of electrons among nearest 
neighboring sites. Thus, one could show that the real space 
Green function $G(r)$ (we drop the frequency label from this 
point on for simplicity) for $r$ nearest neighbor hops is proportional 
to $G(r)\sim t^r$ as $t\rightarrow 0$. On the other hand, Metzner 
{\it et al.,} and M\"uller-Hartmann \cite{Metzner,muller-hartmann} 
have shown that in $D$ dimensions, the factor $t$ should be 
renormalized as $t/\sqrt{D}$ in order to have a finite 
density of states width as $D\rightarrow\infty$. 
As a result of this renormalization  
\begin{eqnarray}
\label{eq:G(r)-Dim}
G(r)\sim t^r \sim (1/\sqrt{D})^r \sim D^{-r/2}=e^{-r/r_s} \nonumber\\
&&\hspace{-6.0cm} r_s=\frac{2}{lnD} \,,
\end{eqnarray}
meaning that $G(r)$ falls off exponentially as a function of $r$.

\begin{figure}
\epsfxsize=2.0in
\hspace*{1.5cm}\epsffile{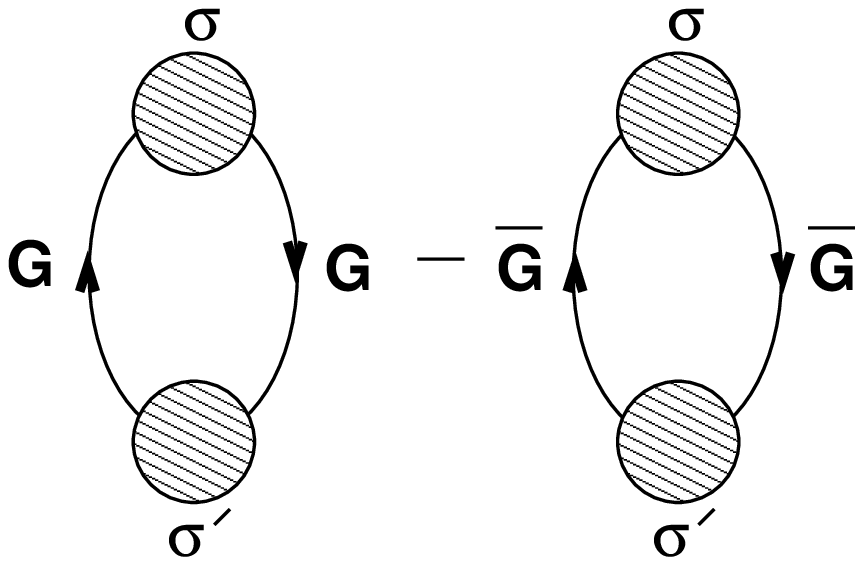}
\caption[a]{\em{First correction by non-compact diagrams,\\ 
$\delta^{(1)}[\Delta\Omega_{ncp}]$.}}
\label{noncomp-diff}
\end{figure}

\begin{figure}
\epsfxsize=2.2in
\hspace{1.2cm}\epsffile{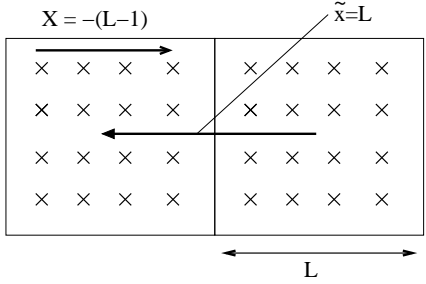}
\caption[a]{\em{Two adjacent clusters with their $\X$ and $\tilde{\x}$ 
vectors.}}
\label{clust_vect}
\end{figure}

\begin{figure}
\epsfxsize=2.7in
\hspace*{1.2cm}\epsffile{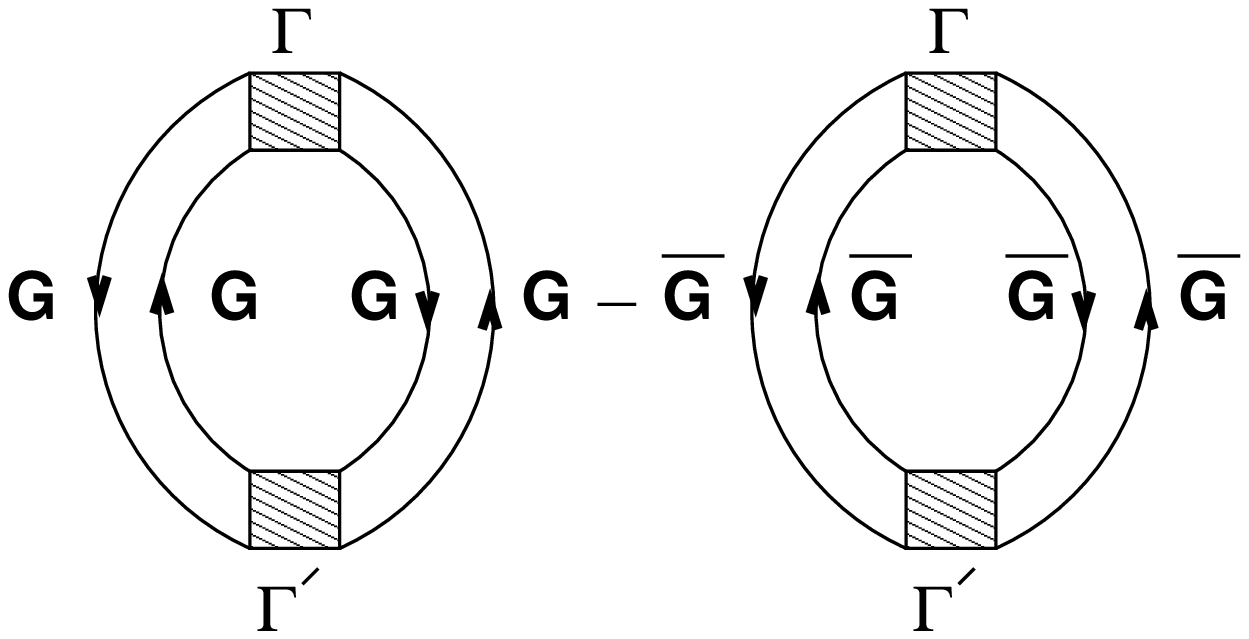} 
\caption[a]{First correction by compact diagrams, 
$\delta^{(1)}[\De\Omega_{cp}]$.
}
\label{comp-diff}
\end{figure}

In the DCA, we attempt to minimize the error due to coarse-graining 
the Green function (and potentials) in the Feynman diagrams. 
Consider the first non-trivial correction to the coarse-grained non-compact 
diagrams generated by replacing the explicit coarse-grained Green function 
lines by the non-coarse-grained ones as illustrated in Fig.~\ref{noncomp-diff}
\begin{eqnarray}
\delta^{(1)}[\Delta\Omega_{ncp}]\sim\frac{1}{N}\sum_{\k}\sigma(\K)
\sigma'(\K)G(\k)^{2}-\frac{1}{N_c}\sum_{\K}\nonumber\\
&&\hspace*{-8.0cm}\sigma(\K)\sigma'(\K)\bar{G}(\K)^{2}\,,
\label{delta_non_comp}
\end{eqnarray}
where $\K$ are the coarse-graining cells momenta and 
$\k=\K+\tilde{\k}$ include all the momenta in the 
first Brillouin zone shown in Fig.~\ref{divide_x_k}. 
In this derivation we also presume that the self-energy is 
$\tilde{\k}$ independent and the entire $\tilde{\k}$ dependence 
is  embedded in the Green functions.
By breaking up the sums over $\k$ in Eq.~\ref{delta_non_comp} 
into $\K$ and $\tilde{\k}$ sums and writing all the G in terms 
of their Fourier transforms at the same time we get
\begin{eqnarray}
\delta^{(1)}[\Delta\Omega_{ncp}]\sim\frac{1}{N}\sum_{\tilde{\k},\K}
\sigma(\K)
\sigma'(\K)\sum_{{\x}_1,{\x}_2}G({\x}_1)G({\x}_2)\times\nonumber\\
&&\hspace*{-8.0cm}e^{-i\tilde{\k}\cdot({\x}_1+{\x}_2)}
e^{-i\K\cdot({\x}_1+{\x}_2)}-\frac{N_c}{N^2}\sum_{\K,\tilde{\k}_1,
\tilde{\k}_2}\sigma(\K)\sigma'(\K)\times\nonumber\\
&&\hspace*{-8.0cm}\sum_{{\x}_1,{\x}_2}G({\x}_1)G({\x}_2)
e^{-i\tilde{\k}_1\cdot{\x}_1}e^{-i\tilde{\k}_2\cdot{\x}_2}
e^{-i\K\cdot({\x}_1+{\x}_2)} \,,
\label{non_comp_FT}
\end{eqnarray}
in which we used Eq.~\ref{eq:cgGDCA} for $\bar{G}(\K)$. According to 
Fig.~\ref{divide_x_k}, we can split $\x=\X+\tilde{\x}$ where $
\tilde{\x}$ extend between two separate clusters while $\X$ always 
remain within a single cluster. By making this separation in 
Eq.~\ref{non_comp_FT} one picks up phases including products of 
$-i\K\cdot\X$, $-i\K\cdot\tilde{\x}$, $-i\tilde{\k}\cdot\X$ and 
$-i\tilde{\k}\cdot\tilde{\x}$ in their exponents. The phase associated 
with the product $\K\cdot\tilde{\x}=2n\pi$ with $n$ an integer 
equals unity. The phases involving $-i\tilde{\k}\cdot\X$ products 
are also neglected as discussed in section~\ref{Dynamical}. 
Hence, Eq.~\ref{non_comp_FT} can be rewritten as follows 
\begin{eqnarray}
\delta^{(1)}[\Delta\Omega_{ncp}]\sim\frac{1}{N}\sum_{\K}\sigma(\K)
\sigma'(\K)\sum_{{\X}_1,{\X}_2}\sum_{\tilde{\x}_1,\tilde{\x}_2}
G({\X}_1+\tilde{\x}_1)\times\nonumber\\
&&\hspace*{-9.0cm}\G({\X}_2 +\tilde{\x}_2)e^{-i\K\cdot({\X}_1+{\X}_2)}
\sum_{\tilde{\k}}e^{-i\tilde{\k}\cdot(\tilde{\x}_1+\tilde{\x}_2)}-
\frac{N_c}{N^2}\sum_{\K}\sigma(\K)\times\nonumber\\
&&\hspace*{-9.0cm}\sigma'(\K)\sum_{{\X}_1,{\X}_2}
\sum_{\tilde{\x}_1,\tilde{\x}_2}G({\X}_1+\tilde{\x}_1)
\G({\X}_2+\tilde{\x}_2)e^{-i\K\cdot({\X}_1+{\X}_2)}
\times\nonumber\\
&&\hspace*{-9.0cm}\sum_{\tilde{\k}_1,\tilde{\k}_2}
e^{-i\tilde{\k}_1\cdot{\x}_1}e^{-i\tilde{\k}_2\cdot{\x}_2} \,.
\label{non_comp_mdf}
\end{eqnarray}
Implementing the following substitutions
\begin{eqnarray}
\sum_{\tilde{\k}}e^{-i\tilde{\k}\cdot(\tilde{\x}_1+\tilde{\x}_2)}=
\frac{N}{N_c}\delta_{\tilde{\x}_1,-\tilde{\x}_2}\hspace*{0.3cm}
\textrm{and}\hspace*{0.3cm}\sum_{\tilde{\k}_1,\tilde{\k}_2}
e^{-i\tilde{\k}_1\cdot{\x}_1}e^{-i\tilde{\k}_2\cdot{\x}_2}\nonumber\\
&&\hspace*{-9.0cm}=\big(\frac{N}{N_c}\big)^2\delta_{\tilde{\x}_1,0}
\delta_{\tilde{\x}_2,0} \,,
\label{kroneker}
\end{eqnarray}
Eq.~\ref{non_comp_mdf} simplifies into
\begin{eqnarray}
\delta^{(1)}[\Delta\Omega_{ncp}]\sim\frac{1}{N_c}\sum_{\K}\sigma(\K)
\sigma'(\K)\big[\sum_{{\X}_1,{\X}_2,\tilde{\x}}G({\X}_1+\tilde{\x})
\times\nonumber\\
&&\hspace*{-8.5cm}\G({\X}_2-\tilde{\x})e^{-i\K\cdot({\X}_1+{\X}_2)}
-\sum_{{\X}_1,{\X}_2}G({\X}_1)\G({\X}_2)\times\nonumber\\
&&\hspace*{-8.5cm}e^{-i\K\cdot({\X}_1+{\X}_2)}\big] \,.
\label{non_comp_smplfy}
\end{eqnarray} 
Setting $\sigma(\K)\sigma'(\K)=\xi(\K)$ and performing the $\K$ summation
\begin{eqnarray}
\delta^{(1)}[\Delta\Omega_{ncp}]\sim\sum_{{\X}_1,{\X}_2}
\xi({\X}_1+{\X}_2)\big[\sum_{\tilde{\x}}G({\X}_1+\tilde{\x})
\times\nonumber\\
&&\hspace*{-8.0cm}\G({\X}_2-\tilde{\x})-G({\X}_1)\G({\X}_2)\big]=
 \sum_{{\X}_1,{\X}_2}\xi({\X}_1+{\X}_2)\times\nonumber\\
&&\hspace*{-8.0cm}\sum_{\tilde{\x}\neq0}G({\X}_1+\tilde{\x})
\G({\X}_2-\tilde{\x})\,.
\label{non_comp_fullyR}
\end{eqnarray}
Knowing that $\xi({\X}_1+{\X}_2)=\sum_{_{\X}}\sigma(\X)
\sigma'({\X}+{\X}_1+{\X}_2)$ and also that the lowest order of 
$\sigma(\X)\sim G^{3}(\X)\sim e^{-3|\X|/r_s}$ we conclude that in 
Eq.~\ref{non_comp_fullyR}, the largest contribution is due to terms 
having ${\X}_1=-{\X}_2$ or in other words, local $\xi$. As shown in 
Fig.~\ref{clust_vect}, the first term in the $\tilde{\x}$ sum 
corresponds to $|\tilde{\x}|=L$ (size of the cluster) and $\X$ can 
be as large as $\X=-(L-1)$ in the opposite direction. 
Hence, the leading order term in Eq.~\ref{non_comp_fullyR} falls off as

\begin{eqnarray}
\delta^{(1)}[\Delta\Omega_{ncp}]\sim\xi(0)\times 2D\times G(L-(L-1))
\times\nonumber\\
&&\hspace*{-7.0cm} G(-L+(L-1))\sim 2D\xi(0)e^{\frac{-1}{r_s}}
e^{\frac{-1}{r_s}}=2D\xi(0)e^{\frac{-2}{r_s}}\,,
\label{non_comp_final}
\end{eqnarray}
where $2D$ is the number of $|\tilde{\x}|=L$ contributions in $D$ 
different dimensions of a $D$ dimensional cubic lattice. In 
Eq.~\ref{non_comp_final} we also used the fact that due to the 
lattice symmetry, $G(-\X)=G(\X)$. As $D\rightarrow\infty$, using 
$r_s=2/lnD$
\begin{eqnarray}
\delta^{(1)}[\Delta\Omega_{ncp}]\sim 2D\xi(0)e^{-lnD}= 
2\xi(0)DD^{-1}\sim\ {\cal O}(1)\,.
\label{non_comp_infty}
\end{eqnarray}
which indicates the existence of non-local corrections to the 
non-compact contribution of the thermodynamic potential even at 
infinite dimensions.
Now we replace the coarse-grained self-energy bubbles in 
Fig.~\ref{noncomp-diff} 
with coarse-grained vertices having four external legs and look at the 
difference between compact diagrams with and without coarse-grained Green functions which are explicitly shown in the figure. Since we earlier dropped the frequency labels in the Green functions, here we use indices $1, 2, 3$ and $4$ to emphasize that these Green functions have different frequency labels. The first correction to the compact contribution of the thermodynamic potential 
depicted in Fig.~\ref{comp-diff} is
\begin{eqnarray}
&& \delta^{(1)}[\De\Omega_{cp}]\sim\frac{1}{N^3}\sum_{\stackrel{\k_1,\k_2}
{_{\q}}}\Gamma(\K_1,\K_2,\Q)\Gamma'(\K_1,\K_2,\Q) G_1(\k_1)
\times\nonumber\\
&& G_2(\k_2)G_3(\q)G_4(\k_1+\k_2-\q)-\frac{1}{N_c^3}
\sum_{\stackrel{\K_1,\K_2}
{_{\Q}}}\Gamma(\K_1,\K_2,\Q)\times\nonumber\\ 
&& \Gamma'(\K_1,\K_2,\Q)\bar{G}_1(\K_1)\bar{G}_2(\K_2)\bar{G}_3(\Q)
\bar{G}_4(\K_1+\K_2-\Q)\,,\nonumber\\
&&  
\label{delta_comp}
\end{eqnarray}
where similar to Eq.~\ref{delta_non_comp}, all the vertices are 
coarse-grained but the Green functions are not.
Following the same procedure as for the non-compact difference we 
arrive at
\begin{eqnarray}
&& \delta^{(1)}[\De\Omega_{cp}]\sim\sum_{\stackrel{\X_1,\X_2}
{_{\X_3,\X_4}}}\Lambda(\X_1+\X_4,\X_2+\X_4,\X_3-\X_4)\times
\nonumber\\
&& \big[\sum_{\tilde{\x}}G_1(\X_1-\tilde{\x})
 G_2(\X_2-\tilde{\x})G_3(\X_3+\tilde{\x})G_4(\X_4+\tilde{\x})-\nonumber\\
&& G_1(\X_1)G_2(\X_2)G_3(\X_3)G_4(\X_4)\big]\,,
\label{delta_comp_fullyR}
\end{eqnarray}
where
\begin{eqnarray}
\Lambda(\X_1+\X_4,\X_2+\X_4,\X_3-\X_4)= \frac{1}{N_c^3}
\sum_{\stackrel{\K_1,\K_2}
{_{\Q}}}\Gamma(\K_1,\K_2,\Q)\times\nonumber\\
&&\hspace{-9.2cm}\Gamma'(\K_1,\K_2,\Q)e^{-i{\K}_1\cdot({\X}_1+{\X}_4)}
e^{-i{\K}_2\cdot({\X}_2+{\X}_4)}e^{-i\Q\cdot({\X}_3-{\X}_4)}=\nonumber\\
&&\hspace{-9.2cm}\sum_{\stackrel{\X,\X'}{_{\X''}}}\Gamma(\X,\X',\X'')
\times\nonumber\\
&&\hspace{-9.2cm}\Gamma'(\X+\X_1+\X_4,\X'+\X_2+\X_4,\X''+\X_3-\X_4)\,.
\label{Lambda}
\end{eqnarray}
Once again, the largest contribution is associated with local
$\Lambda$, i.e., $\X_1=-\X_4$, $\X_2=-\X_4$ and $\X_3=\X_4$. Therefore
\begin{eqnarray}
\delta^{(1)}[\De\Omega_{cp}]\sim\sum_{\X}\Lambda(0)\big[\sum_{\tilde{\x}}
G_1(-\X-\tilde{\x})G_2(-\X-\tilde{\x})\times\nonumber\\
&&\hspace{-8.3cm}G_3(\X+\tilde{\x})G_4(\X+\tilde{\x})-
G_1(\X)G_2(\X)G_3(\X)G_4(\X)\big]=\nonumber\\
&&\hspace{-8.3cm}\Lambda(0)\sum_{\X,\tilde{\x}\neq0}G_1(\X+
\tilde{\x})G_2(\X+\tilde{\x})G_3(\X+\tilde{\x})G_4(\X+\tilde{\x})\,.
\nonumber\\&&\hspace{-8.3cm}
\label{delta_comp_loclamb}
\end{eqnarray}
Considering $|\tilde{\x}|=L$ and $\X_{min}=-(L-1)$
\begin{eqnarray}
\delta^{(1)}[\De\Omega_{cp}]\sim\Lambda(0)\times2D\times
G_1(-L+(L-1))\times\nonumber\\
&&\hspace{-7.5cm}G_2(-L+(L-1))G_3(L-(L-1))G_4(L-(L-1))=\nonumber\\
&&\hspace{-7.5cm}2D\Lambda(0)e^{\frac{-4}{r_S}}=2\Lambda(0)D^{-1}\,,
\hspace{1.0cm}(r_s=\frac{2}{lnD})
\label{delta_comp_exp}
\end{eqnarray}
which vanishes as $D\rightarrow\infty$. 

Comparing Eq.~\ref{delta_comp_exp} with Eq.~\ref{non_comp_infty} shows 
that the first correction to the compact contribution of the 
thermodynamic potential falls off exponentially twice as fast as 
the equivalent correction in the non-compact contribution. 
In addition, even at the infinite dimensional limit, there are 
corrections of order one to the non-compact contribution whereas 
for the compact diagrams the DCA becomes exact and there are no 
corrections. This justifies coarse-graining only in the compact 
diagrams. The Green functions in the non-compact diagrams have to 
be explicitly constructed from the coarse-grained self-energy 
$\Sigma_{DCA}(\K,\omega_n)$ using Eq.~\ref{G_DCA}. 
\vspace{-0.1cm} 
\begin{figure}
\epsfxsize=3.0in
\epsffile{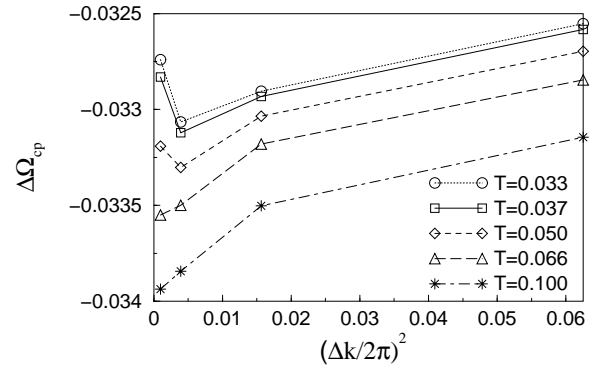} 
\hspace{-2.5in}
\caption[a]{The compact contribution to the thermodynamic potential 
versus $(\Delta k)^2$ at $U/t=1.57$ and various temperatures, using
coarse-grained Green functions. The deviation from linearity at the
lowest temperatures hint at the correlation length exceeding the
cluster size.
}
\label{compact-DCA}
\end{figure}
\vspace{-0.5cm}
\begin{figure}
\epsfxsize=3.0in
\epsffile{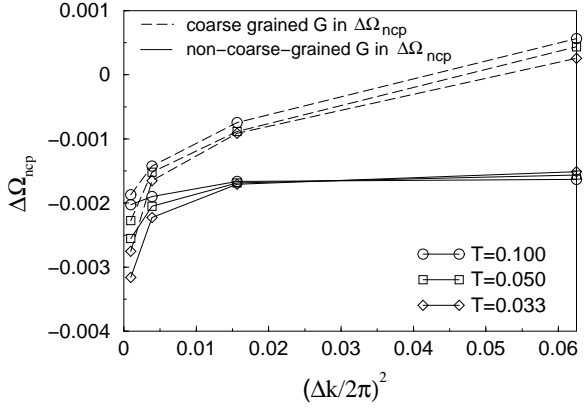} 
\caption[a]{Non-compact contribution to the thermodynamic potential 
versus $(\Delta k)^2$ at $U/t=1.57$ with and without coarse-grained 
Green functions. Using coarse-grained Green functions can result in a 
non-compact contribution with the wrong sign.
}
\label{noncompact-DCA}
\end{figure}
In Fig.~\ref{compact-DCA}, the compact contribution of the
thermodynamic 
potential difference constructed by coarse-grained Green functions is 
plotted versus $(\Delta k)^2=1/L^2$. The variation of 
$\Delta \Omega_{cp}$ over the entire $1/L^2$ range is about 
$1\%$. At very low temperatures ($T<0.066$ in the inset), some 
deviation from linearity is observed due to the correlation length 
exceeding the cluster size $L$ and therefore the approximation of 
$\Sigma(k)$ by $\Sigma_{DCA}(K)$ begins to break down. 
Fig.~\ref{noncompact-DCA} illustrates the non-compact contribution 
sketched versus $1/L^2$ using both coarse-grained and 
non-coarse-grained Green functions. Using non-coarse-grained 
Green functions, the variation of $\Delta \Omega_{ncp}$ over the
entire $1/L^2$ range is roughly $30\%$. 
Coarse-graining the Green functions in these diagrams can even change
the sign of this non-compact contributions,
clearly indicating that 
coarse-graining the Green function is an unrealistic approximation 
for non-compact diagrams. However, one notices that for large 
cluster sizes the coarse-grained and non-coarse-grained results 
approach each other as the approximation to the infinite lattice 
becomes better.
\section{DCA in frequency space}
\label{Freq-DCA}
\begin{figure}
\epsfxsize=3.6in
\hspace{-0.2cm}\epsffile{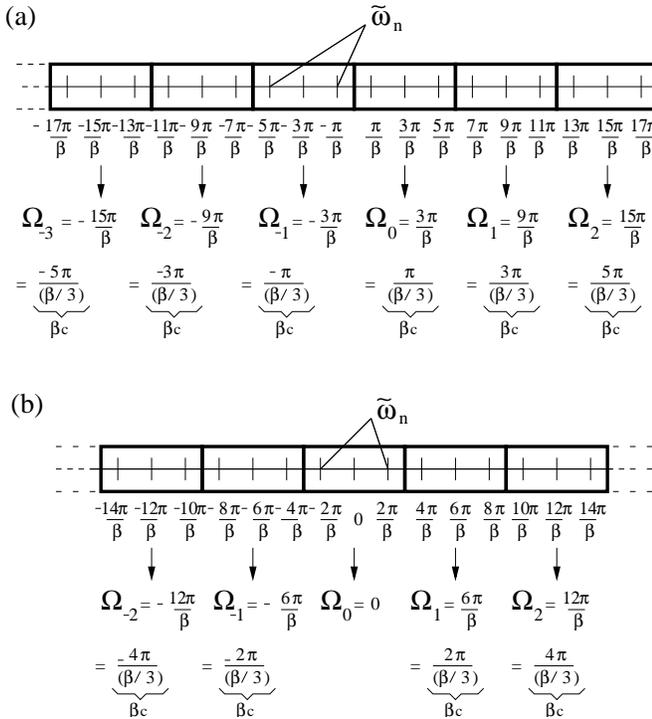} 
\vspace{0.05cm}
\caption[a]{Dividing the fermion (a) and boson (b) Matsubara frequency 
spaces into subcells. The
central $\Omega_n$ frequencies can be written in the form of the 
original lattice with renormalized $\beta$ denoted as $\beta_c$ 
($\beta_c=\beta/3$ in this case).
}
\label{MB-DCA}
\end{figure}

As illustrated for the momentum space, the DCA results in a significant 
reduction of the problem complexity and it is complementary to the 
finite size lattice approach. In analogy to the momentum space, one 
could consider dividing the one dimensional Matsubara frequency space 
into a number of coarse-graining subcells. For both fermions and 
bosons, each cell should include an odd number of frequencies in 
order for the frequencies in the centers of these cells to preserve 
Fermionic or Bosonic properties. Fig.~\ref{MB-DCA} represents how 
the frequency space can be divided into coarse-graining subcells 
each comprising a central $\Omega_n$ frequency and a number of 
coarse-graining $\tilde\omega_n$ lying around it. The central 
$\Omega_n$ frequencies can be rewritten in the form of the original 
lattice with renormalized $\beta$ shown as $\beta_c$ in the figure. 
Similar to the case of the momentum space, we make the following 
transformation for the Laue function  

\begin{equation}
\label{eq:MB-Laue}
\Delta=\beta~\de_{\omega_{n1},\omega_{n2}+\omega_{n3}}
\rightarrow\Delta_{DCA}=\beta_c~\de_{\Omega_{n1},\Omega_{n2}
+\Omega_{n3}}\,,
\end{equation}
for the Matsubara frequencies of the vertex shown in Fig.~\ref{Laue} 
considering frequency dependent interactions in general (in 
condensed matter physics, most of the interactions are indeed 
simultaneous and thus frequency independent). As a result, we 
may again coarse-grain the Green function over the subcell frequencies  
\begin{equation}
\label{eq:cgMBDCA}
\bG(\K,\Omega_n)=\frac{\beta_c}{\beta}\sum_{\tilde\omega_n}
G(\K,\Omega_n + \tilde\omega_n)\,.
\end{equation} 
According to Fig.~\ref{MB-DCA}, the full coarse-graining of the 
Green function amounts to $\beta_c\rightarrow0$ which causes all the 
self-energy Feynman diagrams ordered higher than first (Hartree-Fock 
diagrams) to vanish and consequently we arrive at a fully static problem.

Unfortunately, we can show that coarse-graining over Matsubara 
frequencies can lead to the violation of causality and as a result, 
the DCA is not systematically implementable for the Matsubara frequency 
quantities. The simplest example is the non-interacting Green function 
coarse-grained as follows \cite{Pruschke_priv}
\begin{eqnarray}
\label{eq:NI-MBDCA}
\bG^{(0)}(\K,\Omega_n)=\frac{\beta_c}{\beta}\sum_{\tilde\omega_n}G^{(0)}
(\K,\Omega_n + \tilde\omega_n)=\nonumber\\
&&\hspace{-7.5cm}\frac{\beta_c}{\beta}\sum_{\tilde\omega_n}
\frac{1}{i\Omega_n +i\tilde\omega_n-\epsilon_k+\mu} \,.
\end{eqnarray} 
The retarded Green function is derived by the substitution 
$i\Omega_n\rightarrow\Omega+i\eta$
\begin{equation}
\label{eq:Ret-MBDCA}
\bG^{(0)}_{ret}(\K,\Omega)=\frac{\beta_c}{\beta}\sum_{\tilde\omega_n}
\frac{1}{\Omega+i\eta+i\tilde\omega_n-\epsilon_k+\mu}\,.
\end{equation} 
The theory of analytic functions of a complex variable tells us that 
in order for the Green function to remain retarded in the time space, 
$\bG^{(0)}_{ret}(\K,\Omega)$ must not have any poles in the upper half 
plane of $\Omega$. In Eq.~\ref{eq:Ret-MBDCA}, one could readily create 
poles in the upper half plane for negative ${\tilde\omega}_n$ which 
causes causality violation and consequently unphysical results.
\vspace{-0.1in} 
\vspace{-0.2in}
\begin{figure}
\epsfxsize=3.0in
\epsffile{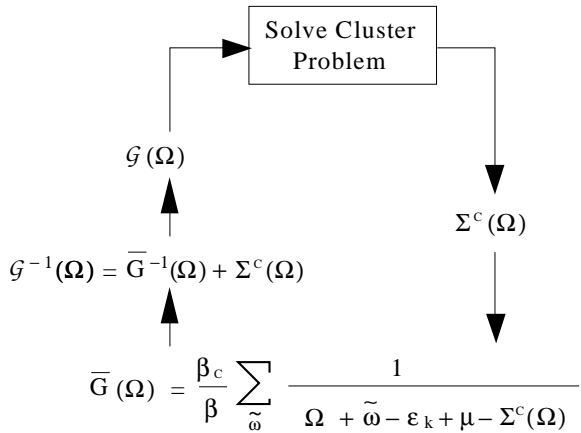} 
\vspace{0.3cm}
\caption[a]{The algorithm for the cluster solvers such as the NCA in 
which at each step before using the cluster solver the bare Green 
function is calculated by excluding the cluster self-energy. This is 
the only step through the algorithm where problems with causality 
might occur. 
}
\label{Cycle-DCA}
\end{figure}

The way around this difficulty is to use real frequencies even at 
finite temperatures. By invoking real frequencies we no longer 
coarse-grain over imaginary values. Therefore, the retarded Green 
function will never acquire poles in the upper half plane and 
remains causal as shown below
\begin{eqnarray}
\label{eq:Ret-GF}
\bG_{ret}(\K,\Omega)=\nonumber\\
&&\hspace{-2.0cm}\frac{N_{\tilde\omega}}{N_\omega}\sum_{\tilde\omega}
\frac{1}{\Omega+\tilde\omega+i\eta+-\epsilon_k
+\mu-\Sigma_{ret}(\K,\Omega)}\,,
\end{eqnarray} 
with $N_\omega$ the total number of frequencies and $N_{\tilde\omega}$ 
the number of those we coarse-graining over in each cell. In an
article by Hettler {\it et al.} \cite{DCA_hettler}, 
a formal proof of causality is given (based on a geometrical argument) 
for coarse-graining in  the momentum space.
This proof can be straightforwardly applied to the real frequency
space as well and it extends the application of real frequency DCA not 
only to the perturbative cluster solvers such as the FLEX but also 
techniques like the NCA.

Lastly, similar to the momentum space, care must be taken when 
choosing the the size of the frequency coarse-graining cells. 
One must make sure that the cells are not larger than some 
characteristic energy scale (e.g. the Kondo temperature $T_K$) 
as the coarse-graining would then suppress the relevant physics. 

\section{Conclusions and Outlook}
\label{Cnclsn}
	We introduce and examine the DCA in detail by employing 
it with the FLEX to study the half filled two dimensional Hubbard 
model. The FLEX is not as precise as nearly exact techniques such 
as quantum Monte Carlo in describing the Hubbard model at strong 
interaction regime.  However, it is capable of illustrating the 
utility of the DCA, including the complementarity and convergence 
of the DCA compared to finite size lattice approaches.  The DCA 
and finite size calculations (with periodic boundary conditions) 
both converge with corrections ${\cal{O}}(\lambda/L^2)$; however, 
in our example the coefficient $\lambda_{DCA}$ was smaller and of 
opposite sign than $\lambda_{FS}$, indicating that the DCA converges 
more quickly and from a complementary direction. This complementarity 
was also seen in other quantities such as the pseudogap in the density 
of state and the non-Fermi liquid behaviour that the DCA (finite size)
calculation systematically under (over) estimates.

	We also provide a detailed microscopic definition of the DCA 
by inspecting the the error generated by coarse-graining the Green 
functions in the compact and non-compact contributions to the thermodynamic 
potential. We conclude that due to the large magnitude of error 
that it generates, coarse-graining the Green function in non-compact 
part should be avoided and only the compact contribution should 
undergo coarse-graining. It also appears that coarse-graining the 
Green functions over the Matsubara frequencies can and will 
lead to the violation of causality and therefore is pathological. 
Nevertheless, one can coarse-grain the Green function over real 
frequencies and preserve the causality not only for the FLEX but 
also cluster solvers such as NCA in which the cluster contribution 
to the coarse-grained dressed Green function is excluded 
before being inserted in the cluster solver.

	The outlook for the FLEX-DCA approach is promising.  Although 
the FLEX fails to accurately describe short-ranged physics such as 
moment formation (and related phenomena like the Mott gap), it
does a good job describing long-ranged physics associated with spin
and charge fluctuations.  On the other hand, numerically exact 
calculations such as QMC, are too expensive to perform for large clusters,
and are thus restricted to the study of short-length scales. However, 
since the DCA gives us a way of parsing the problem into different length 
scales, it may be used to combine the short-length scale information
from the QMC with the long length scale information from the FLEX.  
This may be accomplished, by embedding a QMC cluster, of size $L$, 
into a much larger FLEX cluster of size $L'\gg L$, which is itself 
embedded in a mean-field.  As we have shown here, this approach 
should be implemented by approximating the generating functional 
$\Phi \approx \Phi_{QMC}(L)-\Phi_{FLEX}(L)+\Phi_{FLEX}(L')$.
Work along these lines is in progress.

We would like to acknowledge 
J.J~Deisz, 
D.W.~Hess
Th.~Maier, 
S.~Moukouri, 
and 
Th.~Pruschke 
for very useful discussions and suggestions.  This project was supported 
by NSF Grants No. DMR-0073308 and DMR-9704021.


\begin{references}
%
\bibitem{Pruschke} T.~Pruschke, M.~Jarrell, and J.~K.~Freericks, Adv.\ Phys.\ 
{\bf 42}, 187 (1995).

\bibitem{Georges} A.~Georges, G.~Kotliar, W.~Krauth, and M.~Rozenberg, Rev.\ Mod.\ Phys. {\bf 68}, 13 (1996).

\bibitem{van Dongen} P.~G.~J.~van Dongen, Phys.\ Rev.\ {\bf B 50}, 
14016 (1994).

\bibitem{DCA_hettler} M.~H.~Hettler, M.~Mukherjee, M.~Jarrell, and H.~R.~Krishnamurthy, Phys.\ Rev.\ {\bf B 61}, 12739 (2000).

\bibitem{DCA_hettler2} M.~H.~Hettler, A.~N.~Tahvildar-Zadeh, M.~Jarrell, T.~Pruschke, and H.~R.~Krishnamurthy, Phys.\ Rev.\ {\bf B 58}, 7475 (1998).

\bibitem{DCA_maier1} Th.~A.~Maier, M.~Jarrell, Th.~Pruschke, and J.~Keller, 
Eur.~Phys.~J.~{\bf B 13}, 613-624 (2000).

\bibitem{Bickers1} N.~E.~Bickers, D.~J.~Scalapino, and S.~R.~White, 
Phys.\ Rev.\ Lett. {\bf 62}, 961 (1989).

\bibitem{Bickers2} N.~E.~Bickers, and S.~R.~White, Phys.\ Rev.\ {\bf B 43}, 
8044 (1990).

\bibitem{Deisz} J.~J.~Deisz, D.~W.~Hess, and J.~W.~Serene, Phys.\ Rev.\ Lett.
 {\bf 76}, 1312 (1996) .

\bibitem{Serene} J.W.~Serene, and D.~W.~Hess, Phys.\ Rev.\ {\bf B 44}, 
3391 (1991).

\bibitem{Aryanpour} K.~Aryanpour, M.~H.~Hettler, and M.~Jarrell,
 Phys.\ Rev.\ {\bf B 65}, 153102 (2002). 

\bibitem{Imai} Y.~Imai, and N.~ Kawakami, preprint cond-mat/0204093. 

\bibitem{muller-hartmann} E.~M\"uller-Hartmann, Z.\ Phys.\ {\bf B 74},
507-512 (1989). 

\bibitem{Metzner} W.~Metzner, and D.~Vollhardt, Phys.\ Rev.\ Lett. {\bf 62},
 324 (1989). 

\bibitem{Betts} D.D.~Betts, and G.~E.~Stewart, Can J. Phys. {\bf v 75} n 1
 (1997) p.47-66.

\bibitem{Baym} G.~Baym, Phys.~Rev. {\bf 127}, 1391 (1962).

\bibitem{DCA_maier2} Th.~A.~Maier, and M.~Jarrell, Phys.\ Rev.\ {\bf B 65},
 041104(R) (2002).

\bibitem{Vidberg} H.~J.~Vidberg and J.~W.~Serene, 
J. Low Temp. Phys. {\bf 19}, 
179 (1977).

\bibitem{DHS} J.~J~Deisz, D.~W.~Hess, and J.~W.~Serene, cond-mat/9411026R 
(To appear in "Recent Progress In Many Body Theories", vol. 4, edited by
E. Schachinger, et al. (Plenum, New York)).

\bibitem{Jenkins} M.~A.~Jenkins, and J.~F.~Traub, Comm. ACM 15 (1972) 97-99. 
The routine is available at http://www.netlib.org/tomspdf/419.pdf.

\bibitem{Th-Maier} Th. Maier, private communications.

\bibitem{agd} A.A.\ Abrikosov, L.P.\ Gorkov and  I.E.\ Dzyaloshinski,
{\em{Methods of Quantum Field Theory in Statistical Physics}},
(Dover, New York, 1975).

\bibitem{Pruschke_priv} T.~Pruschke, private communications.
\end{references}
\end{document}